\documentclass[12pt]{article}
\usepackage[dvips]{graphicx}
\usepackage{textcomp}
\usepackage{amssymb}

\begin{document}

\begin{titlepage}
\vskip0.5cm
\begin{flushright}
\end{flushright}
\vskip0.5cm
\begin{center}
{\Large\bf Kosterlitz-Thouless transition in thin films: \\
 A Monte Carlo study of three-dimensional \\
 lattice models} 
\end{center}

\centerline{
\large Martin Hasenbusch
}
\vskip 0.3cm
\centerline{\sl  Institut f\"ur Theoretische Physik, Universit\"at Leipzig
}
\centerline{\sl Postfach 100 920, D-04009 Leipzig, Germany}
\centerline{\sl
e--mail: \hskip 1cm
 Martin.Hasenbusch@itp.uni-leipzig.de}
\vskip 0.4cm
\begin{abstract}
We study the phase transition of thin films in the three-dimensional XY 
universality class.
To this end, we perform a Monte Carlo study of the improved two-component
$\phi^4$ model, the improved dynamically diluted XY model and
the standard XY model on the simple cubic lattice. 
We study films of a thickness up to 
$L_0=32$ lattice spacings. In the short direction of the lattice free boundary 
conditions are employed. Using a finite size scaling (FSS) method, 
proposed recently, we determine the transition temperature 
with high accuracy. The effectively two-dimensional finite size scaling 
behaviour of the Binder 
cumulant $U_4$, the second moment correlation length over the lattice size
$\xi_{2nd}/L$, the ratio of the partition functions with anti-periodic 
and periodic boundary conditions $Z_a/Z_p$ and the helicity 
modulus $\Upsilon$ clearly confirm the Kosterlitz-Thouless 
nature of the transition.  We 
analyse the scaling of the transition temperature with the thickness $L_0$
of the film. The predictions of the renormalization group (RG) theory
are confirmed. We compute the universal ratio of the thickness of the film $L_0$
and the transversal correlation length $\xi_T$ in the three-dimensional 
thermodynamic limit at the 
Kosterlitz-Thouless transition temperature of a film of thickness $L_0$:
$[L_{0,KT}/\xi_T]^* = 1.595(7)$.
This results can be compared with experimental results on thin films of 
$^4$He near the $\lambda$-transition.
\end{abstract}
{\bf Keywords:} $\lambda$-transition, Classical Monte Carlo
simulation, thin films
\end{titlepage}

\section{Introduction}
In the neighbourhood of a second order phase transition the correlation length
diverges as  
\begin{equation}
\label{xipower}
\xi \simeq f_{\pm} |t|^{-\nu} \;,
\end{equation}
where $t=(T-T_c)/T_c$ is the reduced temperature, $f_{+}$ and $f_{-}$ are the amplitudes
in the high and the low temperature phase, respectively, and $\nu$ is the critical 
exponent of the correlation length.
In the neighbourhood of the transition also the behaviour of other quantities is
governed by power laws. E.g. the specific heat behaves as
\begin{equation}
\label{cpower}
C \simeq A_{\pm} |t|^{-\alpha} + B  \;,
\end{equation}
where $B$ is an analytic background, which has
to be taken into account here, since the critical exponent $\alpha$ of the specific heat 
is negative for the three-dimensional XY universality class, as we shall see below.
Critical exponents like $\nu$ and $\alpha$ and ratios of amplitudes such as $f_+/f_-$ 
and $A_{+}/A_{-}$ assume universal values. I.e. they are supposed to assume exactly the
same value for any system in a given universality class.
This can be understood in the framework of the  Renormalization Group (RG).
A universality class is characterised by the dimension of the system, the range
of the interaction and the symmetry of the order parameter. For reviews 
on critical phenomena and the Renormalization Group see e.g.
\cite{WiKo,Fisher74,Fisher98,PeVi02}.

At temperatures below the $\lambda$-transition, $^4$He becomes superfluid.
The order parameter of the $\lambda$-transition
is the phase of a wave function. Therefore it is supposed to share the 
three-dimensional
XY universality class, which is characterized by the O(2), or equivalently U(1),
symmetry of the order parameter. 
The most accurate experimental results for critical exponents and amplitude 
ratios are obtained for the $\lambda$-transition of $^4$He.
E.g. under the condition of micro-gravity, the specific heat has been 
studied for reduced temperatures as small as $t\approx 5 \times 10^{-10}$
\cite{LSNCI-96,lipa2003}, resulting in $\alpha=-0.0127(3)$, corresponding to
$\nu=0.6709(1)$.
Note that the exponents of the specific heat and the correlation length are
related via the hyperscaling relation $d \nu = 2 - \alpha$, where $d$ is the dimension 
of the system.
Experiments on earth, measuring the specific heat and the second sound to
obtain the superfluid density have resulted in accurate estimates of the exponent
of the correlation length $\nu= 0.6717(4)$ and $\nu= 0.6705(6)$ 
in refs. \cite{SiAh84,GoAh92}, respectively. 
For a recent review and an outlook on future experiments in space-stations, 
see ref. \cite{BaHaLiDu07}.  The experimental results for critical
exponents are in reasonable agreement with the most 
precise theoretical prediction for the three-dimensional XY universality class:
$\nu=0.6717(1)$ obtained from a combined Monte Carlo and high temperature 
series study of improved lattice models \cite{recentXY}. For a summary of
theoretical results obtained with various methods (e.g. field-theoretic methods)
see ref. \cite{PeVi02}. Also in the case of universal amplitude ratios, 
there is reasonable match between theoretical results for the 
three-dimensional XY universality class and experimental studies of the 
$\lambda$-transition. See refs. \cite{myAPAM,myamplitude} and refs. therein.

The discussion above refers to the thermodynamic limit. It is a natural 
question how the behaviour at the transition 
is altered by a finite extension of the 
system. It has been addressed both experimentally and theoretically. 
In systems with a finite extension in all directions, there can not be 
any singularity such as eqs.~(\ref{xipower},\ref{cpower}). 
As a remnant of these singularities there remains a peak in the 
neighbourhood of the transition.  With increasing linear extension
the hight of the peak increases and the temperature of the maximum approaches
the critical temperature.

The situation might be different, if the extension in some of the directions
stays infinite. In the case of one infinite direction and two finite ones the 
system becomes effectively one-dimensional. For the type of interactions
that we are dealing with here, there is no phase transition at a finite 
temperature for a one-dimensional system. Hence one expects 
that the behaviour of the effectively one-dimensional system is qualitatively
the same as that of a system which is finite in all directions.

In the case of a thin film geometry, i.e. one direction with 
finite extension and two infinite ones, we expect that the system becomes
effectively two-dimensional in the neighbourhood of the transition. 
Hence, there will be still a phase transition, however it belongs to the 
two-dimensional universality class. I.e. in our case of U(1) symmetry 
of the order parameter, a Kosterlitz-Thouless (KT) transition 
\cite{KT,Jo77,AmGoGr80} is expected.

The behaviour of finite systems in the neighbourhood of a continuous 
transition is described by finite size scaling (FSS). For a reviews see 
\cite{Barber,Privman}.  In essence, close to the transition the bulk 
correlation length $\xi$, i.e. the correlation length in the 
three-dimensional thermodynamic limit, and the extension of the  finite system
are the only relevant length scales. I.e. the physics of finite systems 
is governed by the ratio $L_0/\xi$.
In particular, one expects that, independent on the thickness  
$L_0$ of the thin film, the effectively two-dimensional phase transition 
occurs at a universal value of  $L_0/\xi$. One should note that 
this universal value depends on the type of boundary conditions that are 
applied in the finite direction of the system.
Using eq.~(\ref{xipower}) it follows that \cite{Fi71,CaFi76}
\begin{equation}
\label{TKTscaling}
 T_{c,3d} - T_{KT}(L_0)  \simeq  L_0^{-1/\nu} \;\;,
\end{equation}
where $T_{c,3d}$ is the transition temperature of the three-dimensional 
system and $T_{KT}(L_0)$ the temperature of the KT transition of the 
thin film of thickness $L_0$.

The predictions of finite size scaling have been checked 
by various explicit calculations and experiments \cite{Barber,Privman}. 
For a recent review of experimental results 
near the $\lambda$-transition of $^4$He and $^3$He-$^4$He mixtures
see \cite{GaKiMoDi08}.
For thin films in the three-dimensional XY universality class a multitude 
of field theoretic 
calculations have been performed which allow for comparison with experiment
\cite{Dohm93}. Some important aspects of the behaviour of thin films are 
however not accessible using field theoretic methods. Among these is 
the KT transition, which is the focus of the present work. 

In thin films of $^4$He on various substrates, the order parameter is 
vanishing at the boundary of the film. This has to be taken into account 
in the Monte Carlo simulation of lattice models with film geometry. 
The simplest way to obtain a vanishing field at the boundary are 
so called free boundary conditions. For a precise definition see the 
section below. An alternative are so called staggered boundary conditions.
The numerical results of ref. \cite{NhMa03} confirm that free 
and staggered boundary conditions lead the same results for universal 
quantities. 

On top of the restricted geometry, free boundary conditions might introduce 
new physical effects; For reviews on surface critical phenomena see 
refs. \cite{Binder,Diehl86}. In fact,
free boundary conditions lead to additional corrections to scaling.
The leading one is $\propto L_0^{-1}$  \cite{DiDiEi83}; 
it can be cast in the form $L_{0,eff} = L_0 + L_s$. These corrections
come in addition to those $\propto L_0^{-\omega}$,  
$\propto L_0^{-\omega'}$, ..., which are predicted
for finite systems in general, irrespective of the type of the boundary 
conditions \cite{Barber}. Note that the numerical value of the correction 
exponent is $\omega=0.785(20)$ \cite{recentXY}; similar results are obtained 
with field-theoretic methods; see e.g. the review \cite{PeVi02}. 
The information on $\omega'$ is rather sparse; following ref. \cite{RG}
$\omega' \approx 2 \omega$. 
Fitting Monte Carlo data, it might be difficult to disentangle corrections 
$\propto L_0^{-\omega}$ and  $\propto L_0^{-1}$, leading to sizable errors in 
the extrapolation $L_0 \rightarrow \infty$.
In order to avoid this problem, and to clearly show the existence of 
corrections $\propto L_0^{-1}$ due to the free boundary conditions, we
have studied improved models. In these models the amplitude of 
corrections $\propto L_0^{-\omega}$ vanishes, or in practise, it is so 
small that its effect can be ignored. The precise definition of these
models is given below.

In the literature on can find only a few Monte Carlo studies of lattice models
that address the scaling of  the KT temperature with the thickness of the 
film.
Janke and Nather \cite{JaNa93} have studied the Villain
model on the simple cubic lattice
with free boundary conditions in the short direction. They have determined
the temperature of the KT transition from the behaviour of the correlation 
length
and the magnetic susceptibility in the high temperature phase of the thin films.
They have studied films of a thickness up to $L_0=16$.  Fitting the transition
temperature with eq.~(\ref{TKTscaling}) they find a value for the exponent 
$\nu$ that is about $5\%$ too large compared with the estimates of $\nu$
discussed above.

Schultka and Manousakis \cite{SchMa95a} have studied the standard XY model
with periodic boundary conditions in all directions. They studied lattices 
up to a thickness $L_0=10$. They determined the KT transition temperature
using the effectively two-dimensional 
finite size scaling behaviour of the helicity modulus. They 
conclude that their results are consistent with eq.~(\ref{TKTscaling}), 
using $\nu=0.6705$, which was the best
experimental estimate of the correlation length exponent at the time.
The same authors  \cite{SchMa96,SchMa97} have studied the standard XY model
with staggered boundary conditions in the short direction.  Also in this case,
they have determined the KT transition temperature using the effectively 
two-dimensional finite size scaling behaviour of the helicity modulus.
They studied films of the thicknesses
$L_0=4,8,12,16$ and $20$. They find that their data can be fitted 
with eq.~(\ref{TKTscaling}) using $\nu=0.6705$. However in this case,
that requires that $L_0$ in eq.~(\ref{TKTscaling}) is replaced by
$L_{0,eff} = L_0 + L_s$, with $L_s=5.79(50)$.  They have also reanalysed
the data of ref. \cite{JaNa93} this way. They find that also these results
are compatible with  $\nu=0.6705$, once $L_0$ is 
replaced by $L_{0,eff}$ with $L_s=1.05(2)$. One should note however that 
Schultka and Manousakis do not take into account corrections 
$\propto L_0^{-\omega}$ which should be present in the standard XY
as well as in the Villain model. 

There are further Monte Carlo studies of thin films using lattice models
in the three-dimensional XY universality class.
These studies focus on the specific heat \cite{NhMa03} and refs. therein, 
the thermal resistivity \cite{ZhNhLa06} and refs. therein,
or the thermodynamic Casimir force \cite{Hu07,VaGaMaDi07}. Throughout, the
thickness of the films that had been studied is less or equal to $L_0 =24$.

This paper is organized as follows: 
In the following section, we define the lattice models that we have studied.
Then we introduce the observables that we have measured.
In section \ref{FSSTC} we study the effect of free
boundary conditions in  finite size scaling (FSS) directly at the 
critical temperature of the three dimensional system. 
 In section \ref{MAT}
we briefly summarize the method proposed in ref. \cite{myrecent} to
determine the KT transition temperature. 
Then we determine the KT transition temperature for a large
range of the thickness of the film. We demonstrate that the transition of 
the thin films is indeed of KT nature. To this end we compare the behaviour
of various phenomenological couplings, also called dimensionless quantities in 
the following, at the transition temperature with predictions from
KT theory and from numerical results for two-dimensional systems that are
known to undergo a KT transition.
Next we study the scaling behaviour of the KT transition temperature with
the thickness of the film. Then we compute universal amplitude ratios that
relate the thickness of the film with the correlation length of the bulk
system at the temperature of the KT transition of the thin film. We compare
our estimate for the universal amplitude ratio with that of experiments
on thin films of $^4$He.

\section{The models}
We study various models with $O(2)$-symmetry on a simple cubic lattice.
We consider  systems with film geometry. 
We shall label the sites of the lattice by
$x=(x_0,x_1,x_2)$. The components of $x$ might assume the values 
$x_i \in \{1,2,...,L_i\}$.  We simulate lattices
of the size $L_1=L_2=L$ and $L_0 \ll L$.  In 1 and 2-direction we employ
periodic boundary conditions and free boundary conditions in 0-direction.
This means that the sites with $x_0=1$ and $x_0=L_0$ have only five nearest
neighbours.
This type of boundary condition could also be interpreted as Dirichlet 
boundary conditions with $0$ as value of the field at $x_0=0$ and $x_0=L_0+1$.

The standard XY model is given by the Hamiltonian
\begin{equation}
{\cal H}_{XY} = - \beta \sum_{<x,y>} \vec{s}_x \vec{s}_{y}  \; ,
\end{equation}
where $\vec{s}_x$ is a unit-vector in $\mathbb{R}^2$. $<x,y>$ denotes 
a pair of nearest neighbour sites on the lattice. In our convention, the 
inverse temperature $\beta$ is absorbed into the Hamiltonian.
The Boltzmann factor is given by $\exp(-{\cal H}_{XY})$. The best estimates
of the inverse transition temperature that are quoted in the literature
are $\beta_c=0.454165(4)$, $0.454167(4)$, $0.4541659(10)$ and $0.4541652(11)$
in refs. \cite{spain,Bielefeld2,Bloete05,recentXY}, respectively.
In the following we shall assume $\beta_c=0.4541655(10)$ which is, roughly, the 
average of the estimates given by refs. \cite{Bloete05,recentXY}.

A generalization of the XY model is the $\phi^4$ model on the lattice.
The Hamiltonian is given by
\begin{equation}
{\cal H}_{\phi^4} = - \beta \sum_{<x,y>} \vec{\phi}_x \cdot \vec{\phi}_y
   + \sum_{x} \left[\vec{\phi}_x^2 + \lambda (\vec{\phi}_x^2 -1)^2   \right] \; ,
\end{equation}
where the field variable $\vec{\phi}_x$ is a vector with two real components. 
The partition function is given by
\begin{equation}
Z_{\phi^4} =  \prod_x  \left[\int d \phi_x^{(1)} \,\int d \phi_x^{(2)} \right] \, \exp(-{\cal H}_{\phi^4}).
\end{equation}
In the limit $\lambda \rightarrow \infty$ the field  is fixed to
unit length; i.e. the XY model is recovered. For $\lambda=0$ we get the exactly
solvable Gaussian model.  For $0< \lambda \le \infty$ the model undergoes 
a second order transition that belongs to the XY universality class.
For a discussion see  e.g. \cite{BiJa05}.
Numerically, using Monte Carlo simulations and high-temperature series 
expansions, it has been shown that there is a value $\lambda^* > 0$, where 
leading corrections to scaling vanish.  Numerical estimates of $\lambda^*$
given in the literature are $\lambda^* = 2.10(6)$ \cite{HaTo99}, 
$\lambda^* = 2.07(5)$  \cite{ourXY} and most recently $\lambda^* = 2.15(5)$
\cite{recentXY}.  The inverse of the transition temperature has been 
determined accurately for several values of $\lambda$ using finite size
scaling (FSS) \cite{recentXY}. We shall perform our simulations at 
$\lambda =2.1$, since for this value of $\lambda$ comprehensive Monte 
Carlo studies of the three dimensional system in the low and the 
high temperature phase have been performed \cite{recentXY,myAPAM,myamplitude}.
The inverse critical temperature at $\lambda =2.1$ is $\beta_c=0.5091503(6)$
\cite{recentXY}.
Since  $\lambda =2.1$  is not exactly equal to $\lambda^*$, there are 
still corrections $\propto L^{-\omega}$, although with a small amplitude.
In fact, following ref. \cite{recentXY}, it should be by at least a factor
20 smaller than for the standard XY model.

The dynamically diluted XY model \cite{ourXY,recentXY} is given by
\begin{equation}
{\cal H}_{\rm ddXY} =  -  \beta \sum_{\left<xy\right>}  \vec{\phi}_x \cdot
\vec{\phi}_y - D  \sum_x \vec{\phi}_x^{\,2}
\label{ddxy}
\end{equation}
with the local measure
\begin{equation}
d\mu(\phi_x) =  d \phi_x^{(1)} \, d \phi_x^{(2)} \,
\left[
\delta(\phi_x^{(1)}) \, \delta(\phi_x^{(2)})
 + \frac{1}{2 \pi} \, \delta(1-|\vec{\phi}_x|)
 \right],
 \label{lmeasure}
 \end{equation}
i.e. $|\vec{\phi}_x|$ is either $0$ or $1$, 
 and the partition function
\begin{equation}
Z_{ddXY} = \int \prod_x d\mu(\phi_x)\, \exp(-{\cal H}_{\rm ddXY}) \;\;.
\end{equation}
In the limit $D \rightarrow \infty$ the standard XY model
is recovered. For $D < D_{tri}$ the model undergoes a first order 
transition and for $D > D_{tri}$ a second order transition in the XY 
universality class. 
A mean-field calculation gives $D_{tri}=0$, while from a
improved mean-field calculation we get $D_{tri}<0$ \cite{ourXY}. 
Therefore it is 
quite save to assume that indeed $D_{tri}<0$ for the ddXY model.

From Monte Carlo simulations and high temperature series expansion we 
know that there is a $D_{tri} < D^* < \infty$ such that leading corrections 
to scaling vanish. Numerical estimates are $D^*=1.02(3)$   \cite{ourXY} and
$D^*=1.06(2)$ \cite{recentXY}.  Here we shall present some preliminary 
results for $D=1.02$. At this value of $D$, the inverse 
critical temperature is $\beta_c=0.5637963(4)$.
Also here, the amplitude of leading corrections to scaling
should be by at least a factor 20 smaller than for the standard
XY model  \cite{recentXY}.

\section{The observables}
\label{sdef}
The total magnetisation is defined by
\begin{equation}
 \vec{M} =  \sum_x \vec{\phi}_x \;\;.
\end{equation}
Note that in the case of the XY model we have to replace $\vec{\phi}_x$ by 
$\vec{s}_x$ here and in the following definitions.
The magnetic susceptibility in the high temperature phase, for vanishing 
external field, is given  by
\begin{equation}
\label{chi}
 \chi =  \frac{1}{L_0 L_1 L_2} \langle \vec{M}^2 \rangle \;\;,
\end{equation}
where $ \langle \ldots \rangle$ denotes the expectation value with respect to
the Boltzmann factors defined in the previous section.
The Binder cumulant is defined by
\begin{equation}
 U_4 = \frac{\langle (\vec{M}^2)^2 \rangle}{\langle \vec{M}^2 \rangle^2} \;.
\end{equation}
We also consider the generalization
\begin{equation}
\label{u6}
 U_6 = \frac{\langle (\vec{M}^2)^3 \rangle}{\langle \vec{M}^2 \rangle^3} \;.
\end{equation}

We have computed the
second moment correlation length for the 1- and 2-direction (i.e. the large 
ones).
The second moment correlation length in 1-direction is defined by
\begin{equation}
\xi_{2nd,1} \;=\; \sqrt{\frac{\chi/F_1-1}{4 \; \sin(\pi/L_1)^2}} \;\;\;,
\end{equation}
where
\begin{equation}
F_1 \;= \; \frac{1}{L_0 L_1 L_2} \;   \left \langle
\left |\sum_x \exp\left(i \frac{2 \pi x_1}{L_1} \right) \vec{\phi}_x \right |^2
\right \rangle
\end{equation}
is the Fourier transform of the correlation function at the lowest 
non-zero momentum in 1-direction. In our simulations, we also have  
measured $F_2$ in order to reduce the statistical error.

The helicity modulus $\Upsilon$ gives the reaction of the system under
a torsion. To define the helicity modulus we consider a system, where
rotated boundary conditions are introduced in one direction:
For $x_1=L_1$ and $y_1=1$  the term $\vec{\phi}_x  \vec{\phi}_y$
in the Hamiltonian is replaced  by
\begin{equation}
\vec{\phi}_x \cdot R_{\alpha} \vec{\phi}_y =
\phi_x^{(1)} \left(\cos(\alpha) \phi_y^{(1)} + \sin(\alpha) \phi_y^{(2)} \right)
+\phi_x^{(2)}\left(-\sin(\alpha) \phi_y^{(1)} + \cos(\alpha) \phi_y^{(2)}\right) \;\;.
\end{equation}

The helicity modulus is then given by
\begin{equation}
\label{helidef}
\left . \Upsilon = - 
\frac{L_1}{L_0 L_2}
\frac{\partial^2 \log Z(\alpha)}{\partial \alpha^2} \right|_{\alpha=0} \;\;.
\end{equation}
 Note that we have skipped a factor of $T$ compared with the standard definition
 \cite{FiBaJa73}. Defined this way, the helicity modulus has the dimension of 
 an inverse length. In the literature $\xi_{\perp}=1/\Upsilon$ is referred to as 
transversal correlation length.
The helicity modulus can be directly evaluated in the Monte Carlo simulation.
 Following eq.~(3) of ref. \cite{teli89} one gets for the models considered here
\begin{equation}
\label{helimeasure}
\Upsilon =
\frac{\beta}{L_0 L_1 L_2}
\left \langle \sum_x \vec{\phi}_x \vec{\phi}_{x+(0,1,0)} \right \rangle -
\frac{\beta^2}{L_0 L_1 L_2}
\left \langle  \left[\sum_x (\phi_x^{(1)} \phi_{x+(0,1,0)}^{(2)}
   - \phi_x^{(2)} \phi_{x+(0,1,0)}^{(1)}) \right ]^2  \right \rangle \;\;.
\end{equation}
In addition we have measured the analogous quantity for the $2$-direction.
A problem of the estimator~(\ref{helimeasure}) of $\Upsilon$ is that for a fixed
number of measurements and $L_i \Upsilon$ fixed, the statistical error is increasing
as the critical point is approached; see e.g. ref. \cite{myamplitude}. In section
\ref{algo} below we shall discuss how this problem can be reduced to some extend.
From the definition~(\ref{helidef}) one reads off that $L_0 \Upsilon$ is
dimensionless. Therefore, in the case of thin films the behaviour of $L_0 \Upsilon$  
has to be compared with the prediction for two-dimensional systems in the neighbourhood
of the Kosterlitz-Thouless transition.

The ratio of partition functions
$Z_a/Z_p$ is a useful quantity in the numerical study of phase transitions.
Here $Z_p$ is the partition function of a system with periodic boundary
conditions in two directions of the lattice and free boundary 
conditions in the remaining direction, while $Z_a$ is the partition function
of a system with anti-periodic boundary conditions in one of the directions and
periodic and free boundary conditions in the other two directions. 
Anti-periodic means
that, e.g. in the 1-direction, at the boundary, i.e. for $x_1=L_1$ and $y_1=1$ 
the term
$\vec{\phi}_x  \vec{\phi}_y$
is replaced by
$- \vec{\phi}_x  \vec{\phi}_y$
in the Hamiltonian. This can also be viewed as a rotation by $\pi$. 
The ratio of partition functions can be efficiently measured in 
Monte Carlo simulations using a special version of the cluster algorithm
\cite{Ha92}. The implementation of the algorithm that is used here follows the 
discussion given in Appendix A 2. of ref. \cite{ourXY}.

\section{The Monte Carlo Algorithm}
\label{algo}
For our simulations we use a composition of local and cluster algorithms.
The implementation is similar to that used in ref. \cite{ourXY}. 
See Appendix A 1. of  ref. \cite{ourXY} for details.
Applied to the standard XY model cluster algorithms are ergodic.
However, in the case of the ddXY model and the $\phi^4$ model
additional local updates are needed to change $|\phi_x|$. 
As cluster algorithms, we have used both the single-cluster 
\cite{wolff}  and the  wall-cluster  algorithm \cite{HaPiVi99}.

Let us briefly discuss the details of the local updates for the 
case of the $\phi^4$ model for which most of our simulations were performed.
In the case of the Metropolis update a proposal is generated as
\begin{eqnarray}
 \phi_x^{(1)'} &=& \phi_x^{(1)} + s \left(r- \frac{1}{2} \right) \nonumber \\
 \phi_x^{(2)'} &=& \frac{3}{4} \beta \Phi_x^{(2)}   -  \phi_x^{(2)}
\end{eqnarray}
with $s=3$ (the choice of $s$ is such that very roughly the acceptance rate is
$50 \%$) and 
\begin{equation}
\vec{\Phi}_x = \sum_{y.nn.x}  \vec{\phi}_y \;\;,
\end{equation}
where $y.nn.x$ means that $y$ is a nearest neighbour of $x$. The idea of this
proposal is to change also the second component without using a random number.
The probability to accept this proposal is 
$P_{acc}= \mbox{min}[1,\exp(- \Delta H)]$.  
This update is followed by a second one at the same site, where the role
of the two components is exchanged. Using these updates we run through the 
lattice in typewriter fashion. Going through the lattice once is referred to as
one sweep in the following.

We have also implemented overrelaxation updates for all models that we have simulated.
These are given by
\begin{equation}
 \vec{\phi}_x^{\;\;'} = \vec{\phi}_x
 - 2 \frac{\vec{\Phi}_x \cdot \vec{\phi}_x}{\vec{\Phi}_x^2} \vec{\Phi}_x \;\;,
 \end{equation}
 where
\begin{equation}
\vec{\Phi}_x = \sum_{y.nn.x}  \vec{\phi}_y \;\;.
\end{equation}
Note that these updates do not change the value of the Hamiltonian and therefore no
accept/reject step is needed. Hence it is computationally quite cheap since
no random number and no evaluation of $\exp()$ is needed. Also in the case of the 
overrelaxation update we run through the lattice in typewriter fashion.
As the cluster updates, the overrelaxation update does not change $|\phi_x|$.
The main motivation to use the overrelaxation update is that it allows
to reduce the variance problem of the helicity modulus to some extend.
Since the overrelaxation update is quite cheap in terms of CPU-time, 
we performed several sweeps with the overrelaxation update in each update cycle.
After each of these sweeps the second term on the right hand side of 
eq.~(\ref{helimeasure}) is measured. For a lack of human time we did not
carefully tune the number of overrelaxation sweeps per update cycle.

As random number generator we have used the SIMD-oriented Fast 
Mersenne Twister (SFMT)
\cite{twister} generator.  In particular, we use the {\sl genrand$\_$res53()}
function that produces double-precision output.

\section{Finite size scaling at $\beta_{c,3d}$}
\label{FSSTC}
First we performed simulations  at the transition
temperature of the three-dimensional system. This way we avoid possible 
difficulties related with the KT transition and hence can see more clearly
the effects caused by free boundary conditions. Corrections caused by the
free boundaries should qualitatively not depend on $L_1$ and $L_2$, as
long as $L_1$ and $L_2$ scale with $L_0$. 
Therefore we consider lattices with $L_0 \approx L_1=L_2$, 
allowing us to study a relatively large range in $L_0$.

\begin{figure}
\begin{center}
\scalebox{0.52}
{
\includegraphics{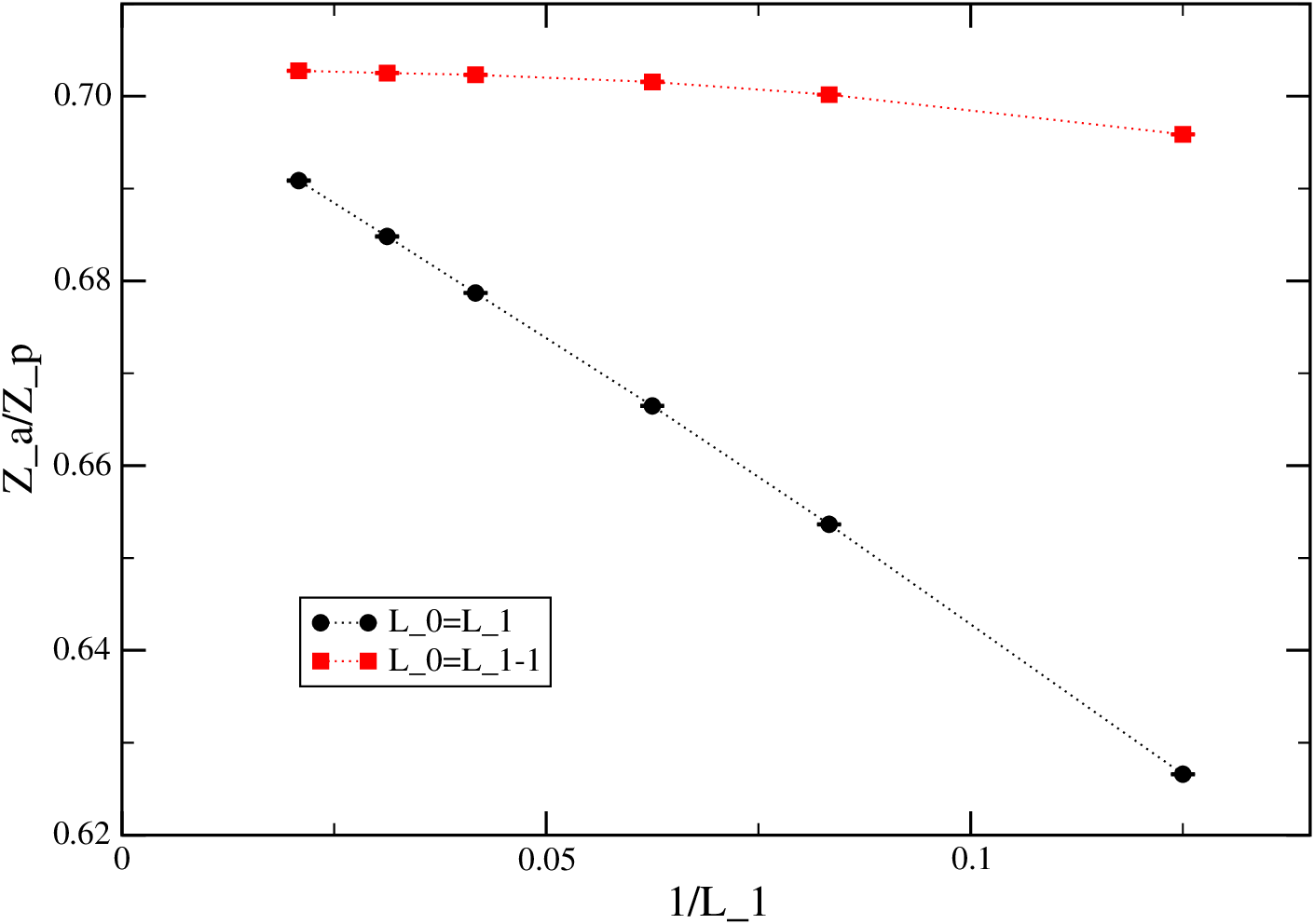}
}
\vskip1.6cm
\scalebox{0.52}
{
\includegraphics{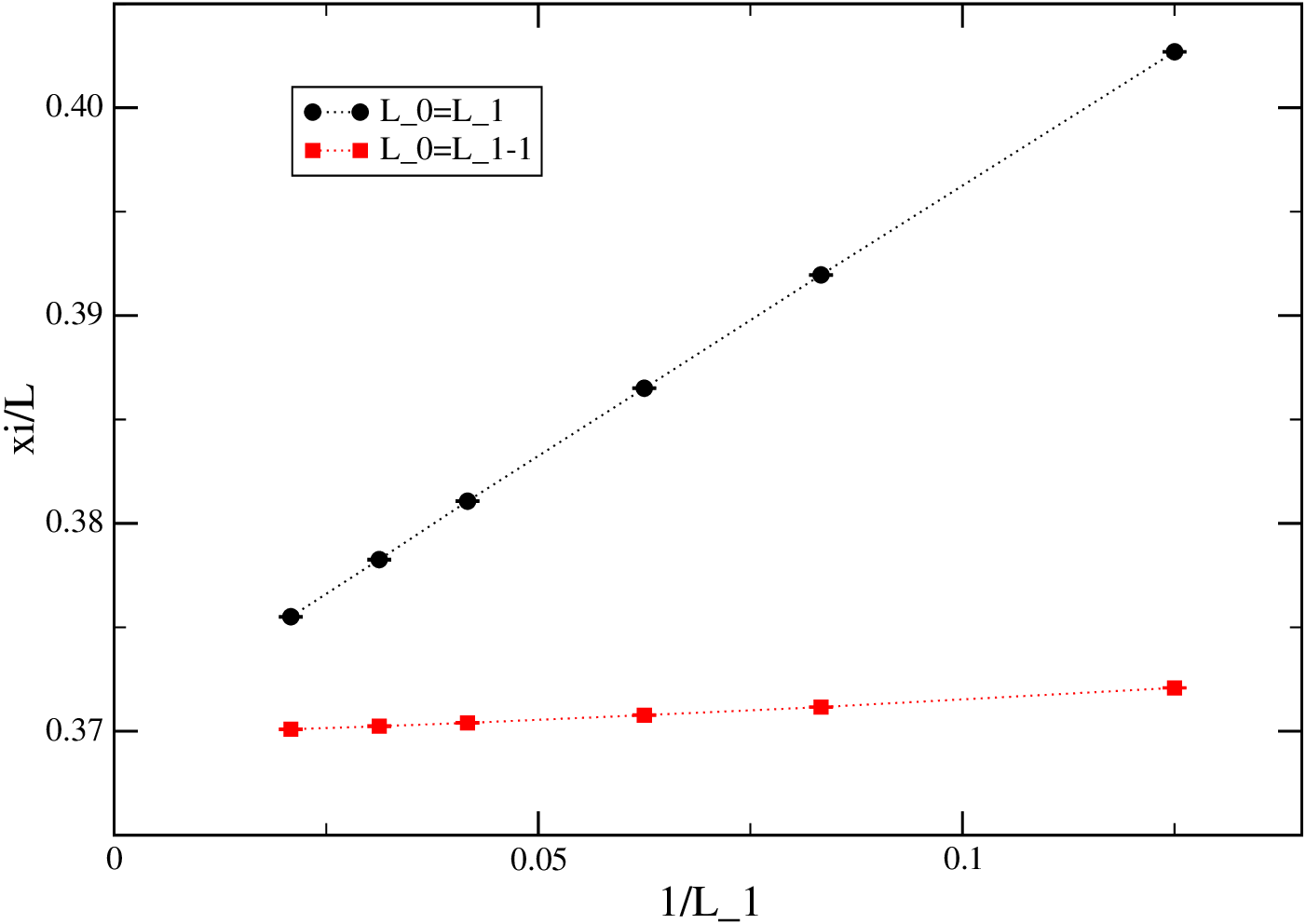}
}
\end{center}
\caption{
\label{FSSplot1} The $\phi^4$ model at $\lambda=2.1$ and $\beta=0.5091503$. 
The upper (lower) figure gives $Z_a/Z_p$ ($\xi_{2nd}/L$) 
as a function of $1/L_1$.
We have simulated systems with  $L_0=L_1=L_2$ and $L_0+1=L_1=L_2$.
}
\end{figure}

\begin{figure}
\begin{center}
\scalebox{0.52}
{
\includegraphics{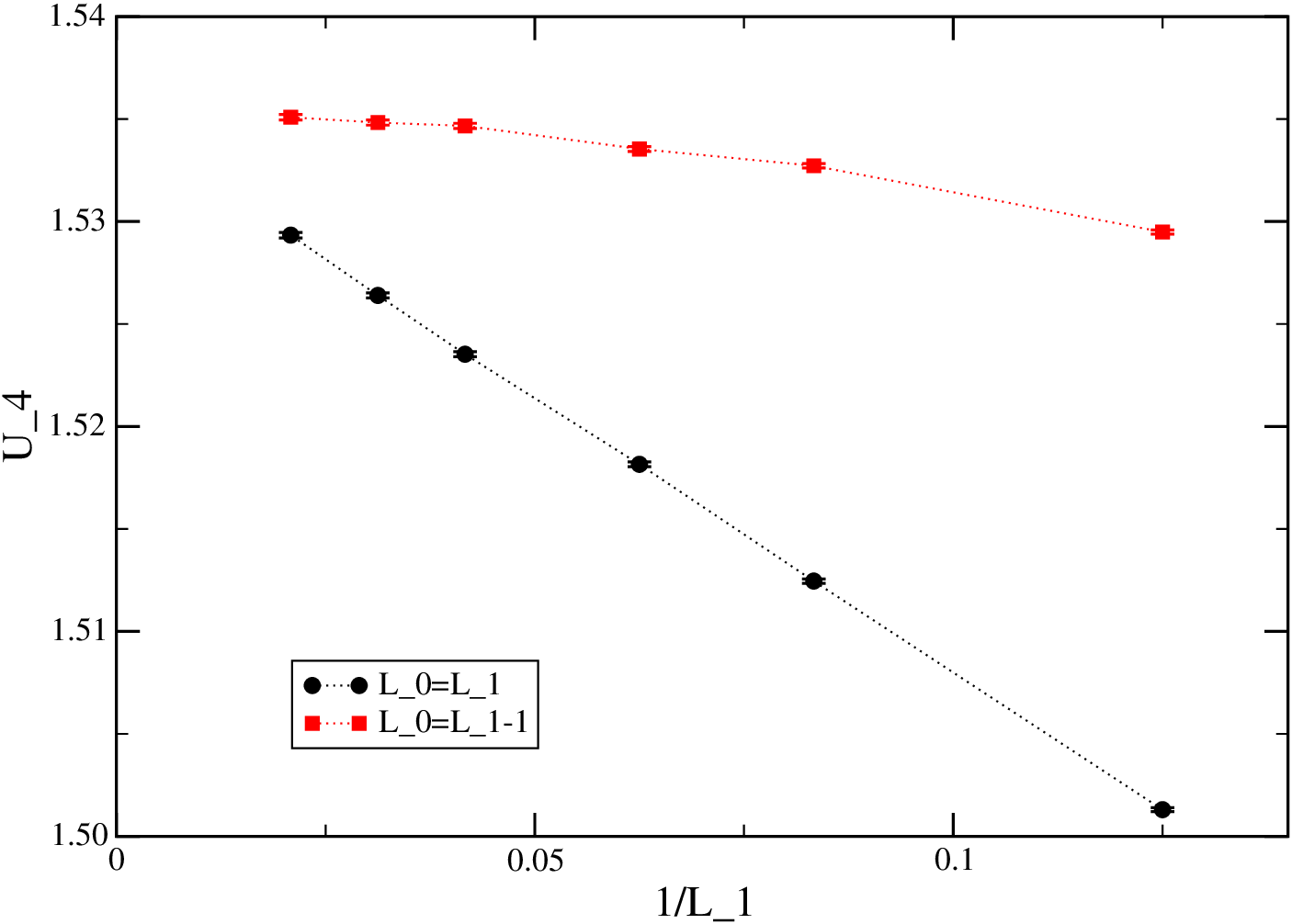}
}
\vskip1.6cm
\scalebox{0.52}
{
\includegraphics{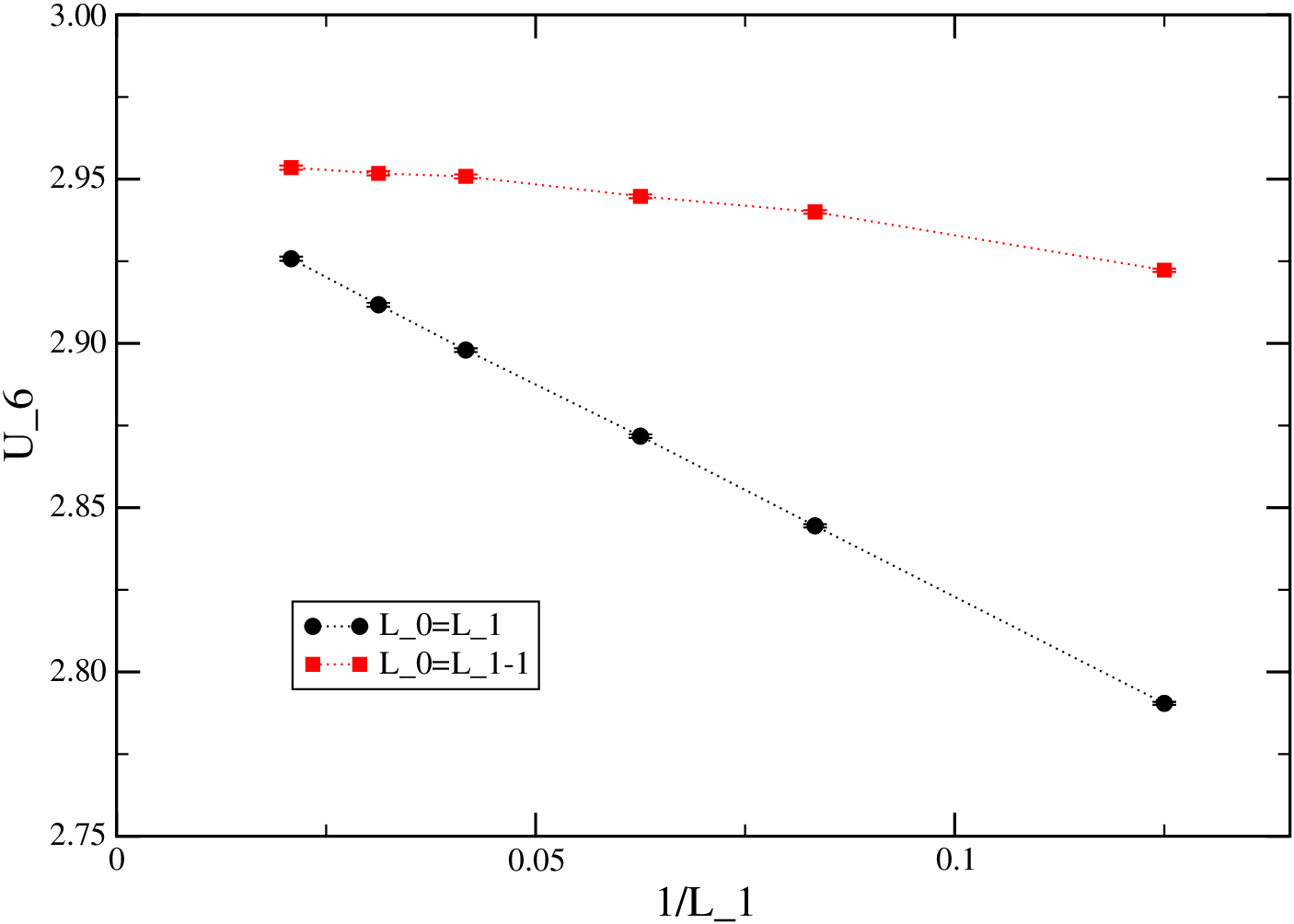}
}
\end{center}
\caption{
\label{FSSplot2}  Same as previous figure, but for
$U_4$ and $U_6$. 
}
\end{figure}

We have simulated the $\phi^4$ model at $\lambda=2.1$ at the best estimate
of the inverse of the critical temperature of the three-dimensional system: 
$\beta=0.5091503$ \cite{recentXY}. We consider the lattice 
sizes $L_1=L_2=8,12,16,24,32$ and $48$.  In addition to $L_0=L_1$ we 
have simulated $L_0=L_1-1$.
Throughout we have performed $10^8$ measurements. For each of 
these measurements, one boundary flip update, 
several single cluster updates, one Metropolis sweep, and three overrelaxation
sweeps were performed. Note that this composition is  
essentially chosen ad hoc.
In total, these simulations took about 6 month of CPU time on one core 
of a Xeon(R) CPU  5160  running at 3 GHz.
We have measured the dimensionless quantities $Z_a/Z_p$, $\xi_{2nd}/L$, 
$U_4$ and $U_6$ as defined in section \ref{sdef}.  
For all these quantities we have determined the coefficients of the 
Taylor series in $\beta$ around $\beta=0.5091503$ up to third order.
This allows us to evaluate these quantities in the neighbourhood of $\beta=0.5091503$.
Our results for $Z_a/Z_p$ and $\xi_{2nd}/L$ are plotted in figure 
\ref{FSSplot1}. The results for $U_4$ and $U_6$ are given in 
figure \ref{FSSplot2}.
In the case of $L_0=L_1$ corrections are clearly visible for all 
four quantities considered.  Plotted as a function of $1/L_1$, 
the data for  $L_0=L_1$ follow roughly a straight line, indicating that
the corrections are proportional to $L_1^{-1}$, as expected. 

In addition, we give results for lattices with $L_0=L_1-1$. Here, for
the largest values of $L_1$, the curves are almost flat for all quantities
considered. E.g. the plots suggest that the dominant corrections can 
be explained by an effective thickness $L_{0,eff} = L_0 + L_s$ with 
$L_s \approx 1$. 

In the following we put this discussion on a more quantitative level 
by performing a sequence of fits of our data for the dimensionless 
quantities.

First we have fitted the dimensionless quantities with the ansatz
\begin{equation}
\label{fssfit1}
 R(L_1,L_1-L_0) = R^* +  c(L_1-L_0) L_1^{-1} \;\;,
\end{equation}
where $R^*$, $c(0)$ and $c(1)$ are the parameters of the fit.
$L_s$ is given by the zero of $c(L_1-L_0)$. Using our data, we have
linearly interpolated/extrapolated $c(0)$ and $c(1)$ resulting in 
\begin{equation}
 L_s = \frac{c(0)}{c(0)-c(1)} \;\;.
\end{equation}

Results of such fits 
are given in the tables \ref{fitzazp1}, \ref{fitxiL1}, \ref{fitU41} and 
\ref{fitU61} for $Z_a/Z_p$, $\xi_{2nd}/L$, $U_4$ and $U_6$, respectively.

\begin{table}
\caption{\sl \label{fitzazp1}
Fits of $Z_a/Z_p$ with the 
ansatz~(\ref{fssfit1}). $L_{1,min}$ is the minimal lattice size 
included into the fit. For a discussion see the text.
}
\begin{center}
\begin{tabular}{|c|l|r|r|}
\hline
\multicolumn{1}{|c}{$L_{1,min}$}&
\multicolumn{1}{|c}{$(Z_a/Z_p)^*$}&
\multicolumn{1}{|c}{$L_s$}&
\multicolumn{1}{|c|}{$\chi^2/$d.o.f.} \\
\hline
\phantom{0}8 &  0.70474(4) &  1.117(1)  & 111.45 \\
12 &  0.70364(5) &  1.068(2)  &  11.84 \\
16 &  0.70326(7) &  1.046(3)  &   1.68 \\
24 &  0.70311(12)&  1.033(7)  &   0.18 \\
\hline
\end{tabular}
\end{center}
\end{table}

\begin{table}
\caption{\sl \label{fitxiL1}
Fits of $\xi_{2nd}/L$ with the
ansatz~(\ref{fssfit1}). $L_{1,min}$ is the minimal lattice size
included into the fit. For a discussion see the text.
}
\begin{center}
\begin{tabular}{|c|l|r|r|}
\hline
\multicolumn{1}{|c}{$L_{1,min}$}&
\multicolumn{1}{|c}{$(\xi_{2nd}/L)^*$}&
\multicolumn{1}{|c}{$L_s$}&
\multicolumn{1}{|c|}{$\chi^2/$d.o.f.} \\
\hline
\phantom{0}8 & 0.36988(2) &  1.066(1) & 16.97  \\
12 & 0.36988(3) &  1.058(2)  &     4.98  \\
16 & 0.36988(4) &  1.053(4)  &     2.80 \\
24 & 0.36986(7) &  1.049(8)  &     0.51 \\
\hline
\end{tabular}
\end{center}
\end{table}

\begin{table}
\caption{\sl \label{fitU41}
Fits of $U_4$ with the
ansatz~(\ref{fssfit1}). $L_{1,min}$ is the minimal lattice size
included into the fit. For a discussion see the text.
}
\begin{center}
\begin{tabular}{|c|l|l|r|}
\hline
\multicolumn{1}{|c}{$L_{1,min}$}&
\multicolumn{1}{|c}{$U_4^*$}&
\multicolumn{1}{|c}{$L_s$}&
\multicolumn{1}{|c|}{$\chi^2/$d.o.f.} \\
\hline
\phantom{0}8 &  1.53578(7)  &   1.188(5)  &   28.39 \\
12 &            1.53545(9)  &   1.119(7)  &    7.57 \\
16 &            1.53540(12) &   1.097(12) &    6.49  \\
24 &            1.53531(20) &   1.055(24) &    0.31  \\
\hline
\end{tabular}
\end{center}
\end{table}

\begin{table}
\caption{\sl \label{fitU61}
Fits of $U_6$ with the
ansatz~(\ref{fssfit1}). $L_{1,min}$ is the minimal lattice size
included into the fit. For a discussion see the text.
}
\begin{center}
\begin{tabular}{|c|l|l|r|}
\hline
\multicolumn{1}{|c}{$L_{1,min}$}&
\multicolumn{1}{|c}{$U_6^*$}&
\multicolumn{1}{|c}{$L_s$}&
\multicolumn{1}{|c|}{$\chi^2/$d.o.f.} \\
\hline
 \phantom{0}8 & 2.9573(3) &  1.224(5) & 33.83 \\
 12 & 2.9556(5) &  1.145(8) &  8.57  \\
 16 & 2.9553(6) &  1.119(12) &   6.84  \\
 24 & 2.9547(10) & 1.070(24) &  0.63 \\
\hline
\end{tabular}
\end{center}
\end{table}

In all four cases, an acceptable $\chi^2$/d.o.f. is only reached for 
$L_{1,min}=24$.  From the fits with $L_{1,min}=24$ we get quite consistent 
results for $L_s$ for the four quantities considered. They range from 
$L_s=1.033(7)$ for $Z_a/Z_p$ up to $L_s=1.070(24)$ for $U_6$. 

In order to check the possible error due to the uncertainty of $\beta_c$, 
we have repeated the fits for $L_{1,min}=24$ for $\beta=0.5091509$.
The result for e.g. $Z_a/Z_p$ is $(Z_a/Z_p)^*=0.70294(12)$ and $L_s=1.0277(7)$.
I.e. the effect is of similar size as the statistical error.

Next we performed fits with the ansatz 
\begin{equation}
\label{fssfit2}
 R(L_1,L_1-L_0) = R^* +  c_1(L_1-L_0) L_1^{-1} +  c_2(L_1-L_0) L_1^{-2} \;\;,
\end{equation}
where now  $R^*$, $c_1(0)$, $c_1(1)$, $c_2(0)$ and $c_2(1)$ are the 
free parameters of the fit.
Here we have not included explicitly $L^{-\omega'}$ corrections, since
at the level of our numerical resolution $\omega'$ is rather close to $2$.

Using this ansatz, we get $\chi^2/$d.o.f. $\approx 1$ already for $L_{1,min}=8$
for all four quantities that we study.
The results for $L_s$ from these fits are slightly smaller than those from 
the ansatz (\ref{fssfit2}): $L_s=0.955(5), 1.040(6), 0.991(19)$ and $0.998(18)$ 
from $Z_a/Z_p$, $\xi_{2nd}/L$, $U_4$ and $U_6$, respectively. In the 
case of $Z_a/Z_p$, where the result for $L_s$ is the smallest, we get 
$L_s=0.969(11)$ from $L_{1,min}=12$. I.e. it is moving towards the other 
results.
Also here, we have checked the effect of the uncertainty in $\beta_c$. 
E.g. for $\beta_c=0.5091509$ and  $L_{1,min}=8$ one obtains
 $L_s=0.950(5)$, $1.035(7)$, $0.981(18)$, and $0.988(18)$, from 
$Z_a/Z_p$, $\xi_{2nd}/L$, $U_4$ and $U_6$, respectively. I.e. the effect 
is of similar size as the statistical error at fixed $\beta$. 

As our final result we quote 
\begin{equation}
\label{LSFSS}
L_s = 1.02(7) 
\end{equation}
where the error-bar covers all the results of the fits reported above.

Here we did not explicitly check the possible error due to residual 
corrections $\propto L^{-\omega}$ at $\lambda=2.1$. 
However, corrections $\propto L^{-\omega}$ should affect 
different phenomenological couplings in a different way. Therefore the 
variation of $L_s$ obtained from different quantities should also  provide
an estimate of the error resulting from residual corrections 
$\propto L^{-\omega}$.

Our final values for the fixed point values of the dimensionless quantities
are $(Z_a/Z_p)^*= 0.7030(5)$, $(\xi_{2nd}/L)^*=0.36986(10)$, $U_4^*=1.5352(2)$, 
and $U_6^*=2.954(1)$.  These values can be compared with those of a $L^3$
system with periodic boundary conditions in all three directions
\cite{recentXY}:
$(Z_a/Z_p)^*=0.3203(1)[3]$, $(\xi_{2nd}/L)^*=0.5924(1)[3]$, $U_4^*=1.2431(1)[1]$,
and $U_6=1.7509(2)[7]$, where the number in $[]$ gives the systematic error due to 
scaling corrections.
The effect of the boundary conditions is very well visible. In the case of 
$(Z_a/Z_p)^*$ the value even differs by more than a factor of two.

The numbers quoted here, could be used to check whether other boundary 
conditions, e.g. staggered boundary conditions, are equivalent with 
free boundary conditions. One would expect a different value of 
$L_s$ but the same values of $R^*$ for equivalent boundary conditions.

\section{Finite size scaling at the KT transition}
\label{MAT}
Finite size scaling is an efficient method to locate and to verify the nature 
of the KT transition in two-dimensional systems.  To this end, usually 
the helicity modulus $\Upsilon$ is studied.  Following ref. \cite{WeMi88} it behaves as
\begin{equation}
\label{ups}
\Upsilon=\frac{2}{\pi} + \frac{1}{\pi} \frac{1}{ (\ln L + C)} \ldots 
\end{equation}
at the transition temperature. 
Recently, we have pointed out \cite{myxi} that for 
lattices with periodic boundary conditions there are numerically small contributions
from winding configurations that are not taken into account in eq.~(\ref{ups}).
Including these contributions one gets \cite{myxi}
\begin{equation}
\label{centralheli}
\Upsilon_{L^2,transition} = 0.63650817819... + \frac{0.318899454...}
{\ln L + C} + \;...
\end{equation}
for an $L^2$ lattice with periodic boundary conditions in both directions
at the KT transition.  
In the appendix we derive  for the ratio of partition functions
\begin{equation}
\label{zazpKT0}
(Z_a/Z_p)_{L^2,transition} = 0.0864272337... - \frac{0.135755793...}{\ln L +C}
 + \;...
\end{equation}
for an $L^2$ lattice at the KT transition. 
We have also determined the leading finite size behaviour of the 
second moment correlation length over the lattice size \cite{myxi}
\begin{equation}
\label{xiexact}
\left . \frac{\xi_{2nd}}{L} \right |_{ L^2,transition} = 0.7506912... + \frac{0.212430...}{\ln L + C}+ \;...
\end{equation}
and more recently for the Binder cumulant \cite{myrecent}
\begin{equation}
\label{binderleading}
U_{4,L^2,transition} = 1.018192(6) -  \frac{0.017922(5)}{\ln L + C}  + ... \;.
\end{equation}
Eqs.~(\ref{centralheli}, \ref{xiexact}) describe quite well the behaviour of numerical 
data for $\Upsilon$ and $\xi_{2nd}/L$  for various models down to rather small 
lattice sizes.
In contrast to that, on the accessible lattice sizes, $U_4$ is decreasing 
with increasing $L$, while eq.~(\ref{binderleading}) suggests that it 
should increase.
This behaviour is explained  by disorder caused
by vortex pairs at a distance of the order $L/2$.  Therefore  a term 
$c_2/(\ln L + C)^2$ has been added to eq.~(\ref{binderleading}).
The coefficient $c_2$ and $C$ have been determined from fits to Monte Carlo
data.  The result for $c_2$ is consistent for different models. This 
confirms the assumption that also the subleading correction in the 
Binder cumulant is universal \cite{myrecent}.

In ref. \cite{myrecent} we suggest that the KT transition of a model can be 
accurately determined by matching $\xi_{2nd}/L$ and the Binder cumulant $U_4$ with 
that obtained from a model where $\beta_{KT}$ is accurately known. In 
ref. \cite{myrecent} we have studied most accurately the dual of the 
absolute value solid-on-solid (ASOS) model. 
In this model, the KT transition occurs at $\tilde \beta=0.80608(2)$
\cite{HaPi97}.

The data of the dual of the ASOS model for $L \ge 96$ are well described by
\begin{eqnarray}
\label{Amatch80608}
\tilde \beta=0.80608 &:& \nonumber \\
U_{4,ASOS}(L) &=& 1.018192 -
  \frac{0.017922}{\ln L -1.18} + \frac{0.06769}{(\ln L -1.18)^2} \nonumber \\
 \frac{\xi_{2nd,ASOS}(L)}{L} &=& 0.7506912 + \frac{0.212430}{\ln L + 0.573}
 \;.
\end{eqnarray}
To check the error due to the uncertainty of $\tilde \beta_{KT}$ one should
repeat the matching with 
\begin{eqnarray}
\label{Amatch80606}
\tilde \beta=0.80606 &:& \nonumber \\
U_{4,ASOS}(L) &=& 1.018192 -
\frac{0.017922}{\ln L -1.193} + \frac{0.06730}{(\ln L -1.193)^2} \nonumber \\
\frac{\xi_{2nd,ASOS}(L)}{L} &=& 0.7506912 + \frac{0.212430}{\ln L + 0.557}
\;.
\end{eqnarray}
In order to match the thin films with these expressions one has to 
adjust $\beta$ and a scale factor for the lattice size.

\section{KT transition in thin films}
\label{KTphi4}
Here we have studied the KT transition of thin films of the $\phi^4$ model at 
$\lambda=2.1$ using the matching method discussed above. We have 
simulated lattices of the thickness $L_0=4$, $6$, $8$, $12$, $16$, $24$ and $32$.
In order to get the phenomenological couplings as  functions of $\beta$ we
have computed the Taylor series of these quantities around $\beta_s$ 
up to the third order, where in the simulation the configurations are generated
with a Boltzmann  factor corresponding to $\beta_s$.
Since the Taylor expansion provides accurate results
only in a small neighbourhood of $\beta_s$, we tried to chose 
$\beta_s \approx \beta_{KT}(L_1)$, where $\beta_{KT}(L_1)$ is the solution of 
eqs.~(\ref{Amatch80608}) for the given lattice size $L_1$. To this end, we 
performed some preliminary simulations or we extrapolated the result for 
$\beta_{KT}(L_1)$ from small lattice sizes $L_1$ to large ones. 
We carefully checked that in 
our final simulations the differences $\beta_{KT}(L_1)-\beta_s$ are small enough 
to ensure that the Taylor series evaluated at $\beta_{KT}(L_1)$ are sufficiently 
accurate for our purpose.

In our final simulations, 
mostly, we performed $2 \times 10^6$ measurements. Only in the case
our largest lattice, $32 \times 1024^2$, we  only measured $1.4 \times 10^6$ times.
For each measurement, 
we performed single cluster updates, boundary flip updates and 3 overrelaxation
sweeps and one Metropolis sweep. The number of single cluster updates was 
chosen such that, on average, the lattice is covered once by the clusters.

E.g. for the $32 \times 1024^2$ lattice, the integrated autocorrelation time 
of the magnetic susceptibility is  $\tau_{\chi} \approx 0.7$. I.e. individual 
measurements are almost statistically independent.
The total CPU time used for these simulations was roughly the equivalent 
of 9 years on a single core of a 2.6 GHz Opteron CPU.
Our results obtained from the matching with the dual
of the ASOS model at $\tilde \beta=0.80608$ are summarized in table 
\ref{matching8to32}.  

\begin{table}
\caption{\sl \label{matching8to32}
Matching the thin film with
the dual of the ASOS model at $\tilde \beta=0.80608$, eqs.~(\ref{Amatch80608}).
The parameters of the matching are the inverse KT temperature $\beta_{KT}$ 
of the thin film and the lattice size $L_{ASOS}$ of the dual of the ASOS model.
}
\begin{center}
\begin{tabular}{|r|r|l|l|}
\hline
\multicolumn{1}{|c}{$L_{0}$}&
\multicolumn{1}{|c}{$L_{1}$}&
\multicolumn{1}{|c}{$\beta_{KT}$}&
\multicolumn{1}{|c|}{$L_{ASOS}$} \\
 \hline
 4 & 32 &   0.610518(24)  &\phantom{0}27.91(8) \\
 4 & 64 &   0.609984(18)  &\phantom{0}53.9(3) \\
 4 &128 &   0.609784(17)  &   113.9(1.3) \\
 4 &256 &   0.609688(12)  &   227.9(3.3) \\
 4 &512 &   0.609699(11)  &   469.(12.) \\
 \hline
 6 &  48&   0.568772(14) &\phantom{0}30.38(10) \\
 6 &  96&   0.568453(12) &\phantom{0}60.37(36) \\
 6 & 192&   0.568313(8)  &  125.1(1.4) \\
 6 & 384&   0.568262(7)  &  260.4(4.6) \\
 6 & 768&   0.568259(7)  &  544.(18.) \\
 \hline
 8 & 64   &   0.549585(9)  &  \phantom{0}31.76(12) \\
 8 & 128  &   0.549381(6)  &  \phantom{0}63.6(4) \\
 8 & 256  &   0.549314(5)  &  133.1(1.5) \\
 8 & 512  &   0.549283(5)  &  272.8(5.5) \\
\hline
12 & 96  &    0.5322567(52)  &  \phantom{0}33.06(12) \\
12 & 192  &    0.5321416(48)  &  \phantom{0}65.9(6) \\
12 & 384  &    0.5321010(33)  &  139.6(1.8) \\
12 & 768  &    0.5320864(28)  &  297.6(6.6) \\
\hline
 16 & 128 &  0.5245609(34) &   \phantom{0}33.87(13) \\
 16 & 256 &  0.5244956(27) &   \phantom{0}67.5(5) \\
 16 & 512 &  0.5244626(23) &  140.5(1.8) \\
 16 & 1024 &  0.5244547(19)&  296.9(6.1) \\
\hline
 24 & 192  &    0.5177878(19)  & \phantom{0}34.39(13) \\
 24 & 384  &    0.5177446(15)  & \phantom{0}68.48(48) \\
 24 & 768  &    0.5177340(11)  &   143.6(1.6) \\
\hline
 32 &  256  &    0.5148522(12) &   \phantom{0}34.88(13) \\      
 32 &  512  &    0.5148259(9) &    \phantom{0}70.4(4) \\  
 32 & 1024  &    0.5148147(9) &    147.7(2.2) \\  
\hline
\end{tabular}
\end{center}
\end{table}

For all values of $L_0$, $\beta_{KT}(L_1)$ is decreasing with increasing 
$L_1$. Apparently the results are converging; The difference 
$\beta_{KT}(L_1)-\beta_{KT}(2 L_1)$ is decreasing with increasing $L_1$.
For $L_0=4$ and $6$, where we have the 
largest range of lattice sizes available, the differences of the results
for $L_1/L_0=64$ and $L_1/L_0=128$ are already smaller than the error-bars.

Eqs.~(\ref{Amatch80608}) are supposed to give the universal 
behaviour within the numerical precision that has been reached in 
ref.~\cite{myrecent}.  Corrections to the universal behaviour are expected to
decay with a power of $L_1$, possibly multiplied by some power of $\ln L_1$.  
There are certainly corrections with 
an exponent $\epsilon = 7/4$ due to the analytic background in the magnetic
susceptibility.  Furthermore there are corrections due to irrelevant scaling 
fields.

In order to check these expectations, we have fitted our estimates of $\beta_{KT}(L_1)$
for $L_0=4$ and $6$, where we have the largest values of $L_1/L_0$ available, with 
the ansatz
\begin{equation}
\label{exbetaKT}
 \beta_{KT}(L_1) =  \beta_{KT}(\infty)  + c L_1^{-\epsilon} \;\;,
 \end{equation}
where $\beta_{KT}(\infty)$, $c$ and $\epsilon$ are the free parameters of the fit.
The results of these fits are summarized in table \ref{fit4and6}.
\begin{table}
\caption{\sl \label{fit4and6}
Results of fits with the ansatz~(\ref{exbetaKT}). $L_{1,min}$ is the 
smallest lattice size that has been included into the fit. For a discussion
see the text.
}
\begin{center}
\begin{tabular}{|c|c|l|l|l|c|}
\hline
\multicolumn{1}{|c}{$L_{0}$}&
\multicolumn{1}{|c}{$L_{1,min}$}&
\multicolumn{1}{|c}{$\beta_{KT}(\infty)$}&
\multicolumn{1}{|c}{$c$}  &
\multicolumn{1}{|c}{$\epsilon$} & 
\multicolumn{1}{|c|}{$\chi^2/$d.o.f.} \\
\hline
4 & 32 & 0.609669(11) & 0.14(4) & 1.47(9) & 2.78 \\
4 & 64 & 0.609681(13) & 0.6(6) & 1.8(3)   & 4.31  \\
\hline
6 & 48  & 0.568241(8) & 0.12(4) & 1.40(9) & 1.48\\
6 & 96  & 0.568250(8) & 0.6(8)  & 1.8(3)  & 0.94\\
\hline
\end{tabular}
\end{center}
\end{table}
The numerical results for the correction exponent $\epsilon$ are not very precise;
they are essentially consistent with a leading correction exponent 
$\epsilon=7/4$ due to the analytic background in the magnetic susceptibility. 

In order to check the uncertainty of our results for $\beta_{KT}$ caused
by the uncertainty of 
the estimate of the KT transition temperature in the ASOS model, we have 
repeated the matching  with
eqs.~(\ref{Amatch80606}). As examples we give the results obtain for 
$L_0=4$ and $L_0=6$ in table \ref{matching080606}.
\begin{table}
\caption{\sl \label{matching080606}
Matching the thin film with
the dual of the ASOS model at $\tilde \beta=0.80606$, eqs.~(\ref{Amatch80606}).
As examples we give the results for $L_0=4$ and $6$.
}
\begin{center}
\begin{tabular}{|r|r|l|l|}
\hline
\multicolumn{1}{|c}{$L_{0}$}&
\multicolumn{1}{|c}{$L_{1}$}&
\multicolumn{1}{|c}{$\beta_{KT}$}&
\multicolumn{1}{|c|}{$L_{ASOS}$} \\
 \hline
 4 &  32 & 0.610532(24) &\phantom{0}28.08(9) \\ 
 4 &  64 & 0.609996(18) &\phantom{0}54.1(3) \\  
 4 & 128 & 0.609793(17) &  114.0(1.3) \\ 
 4 & 256 & 0.609695(12) &  227.7(3.2) \\  
 4 & 512 & 0.609705(11) &  467.(12.) \\
\hline
  6 & 48 & 0.568780(14) &\phantom{0}30.57(10) \\
  6 & 96 & 0.568459(12) &\phantom{0}60.59(36) \\
  6 &192 & 0.568318(8)  & 125.3(1.4) \\
  6 &384 & 0.568266(7)  & 260.1(4.6) \\
  6 &768 & 0.568262(7)  & 542.(18.) \\
\hline
\end{tabular}
\end{center}
\end{table}

We find that the difference in $\beta_{KT}$ compared with the matching
with eqs.~(\ref{Amatch80608}) is smaller than the statistical error
of  $\beta_{KT}$. Also in the case of the matching lattice size $L_{ASOS}$
of the ASOS model the difference is small. These observations also hold for 
larger values of $L_0$ not given in table \ref{matching080606}.

As our  final estimate of the inverse KT transition temperature for 
$L_0=4$ we take $\beta_{KT} =  0.60968(1)[1]$, which is the result of the 
fit with the ansatz~(\ref{exbetaKT}) using $L_{1,min}=64$. 
The number given in $[]$ is an 
estimate of the systematic error. It is estimated from the difference 
of the fit with the ansatz~(\ref{exbetaKT}) using $L_{1,min}=32$ 
and  $L_{1,min}=64$.
Furthermore we take into account the difference between the matching with 
eqs.~(\ref{Amatch80608}) and eqs.~(\ref{Amatch80606}).
In a similar way we arrive at $\beta_{KT} =0.56825(1)[1]$ for $L_0=6$.

For $L_0 \ge 8$  the range of $L_1/L_0$ that we have simulated 
is smaller than for $L_0=4$ and $6$. Therefore, for these values of $L_0$,
we  use $\epsilon=7/4$ fixed in our fits~(\ref{exbetaKT}) 
to obtain $\beta_{KT}$ in the limit $L_1/L_0 \rightarrow \infty$.
In all cases we did not include the smallest lattice size, $L_1/L_0=8$, 
into the fit.
Even if $\epsilon=7/4$ is the dominant correction, subleading corrections
will cause systematic errors. To estimate these, we assume that the effective
correction exponent might take a value in the range  $1.5 < \epsilon < 2$
as it is suggest by our fits for $L_0=4$ and $6$.  To this end, fitting 
with the ansatz~(\ref{exbetaKT}) we have used $\epsilon=1.5$ in addition to
$\epsilon=1.75$. The systematic error is then estimated by the difference of 
these two fits. As a further check of systematic errors, 
we have repeated these fits for the estimates of  $\beta_{KT}$ 
obtained from the matching with the dual of the ASOS model at 
$\tilde \beta=0.80606$, eqs.~(\ref{Amatch80606}).
Our final results are summarized in table \ref{betaKTfinal}.

\begin{table}
\caption{\sl \label{betaKTfinal}
Final results for the inverse KT transition temperature  $\beta_{KT}$
of thin films of the thickness $L_0$. The statistical error is
given in $()$ while the systematic one is quoted in $[]$. For a discussion 
see the text. In addition we give in the last row the inverse critical 
temperature of the three dimensional system.
}
\begin{center}
\begin{tabular}{|r|l|}
\hline
\multicolumn{1}{|c}{$L_{0}$}&
\multicolumn{1}{|c|}{$\beta_{KT}$} \\
\hline
    4    &  0.60968(1)[1] \\
    6    &  0.56825(1)[1] \\
    8    &  0.549278(5)[9] \\
   12    &  0.532082(3)[5] \\
   16    &  0.524450(2)[3] \\
   24    &  0.517730(2)[2] \\
   32    &  0.514810(1)[2] \\
\hline
$\infty$ & 0.5091503(6) \\
\hline
\end{tabular}
\end{center}
\end{table}

From the matching of the thin films with the dual of the ASOS model, 
we obtain also a scale factor that relates the lattice size of the 
thin films and the  dual of the ASOS model. Following the 
RG prediction, thin films of different thickness $L_0$ match, up to 
corrections to scaling, for the same values of $L_1/(L_0+L_s)$.

\begin{figure}
\begin{center}
\scalebox{0.52}
{
\includegraphics{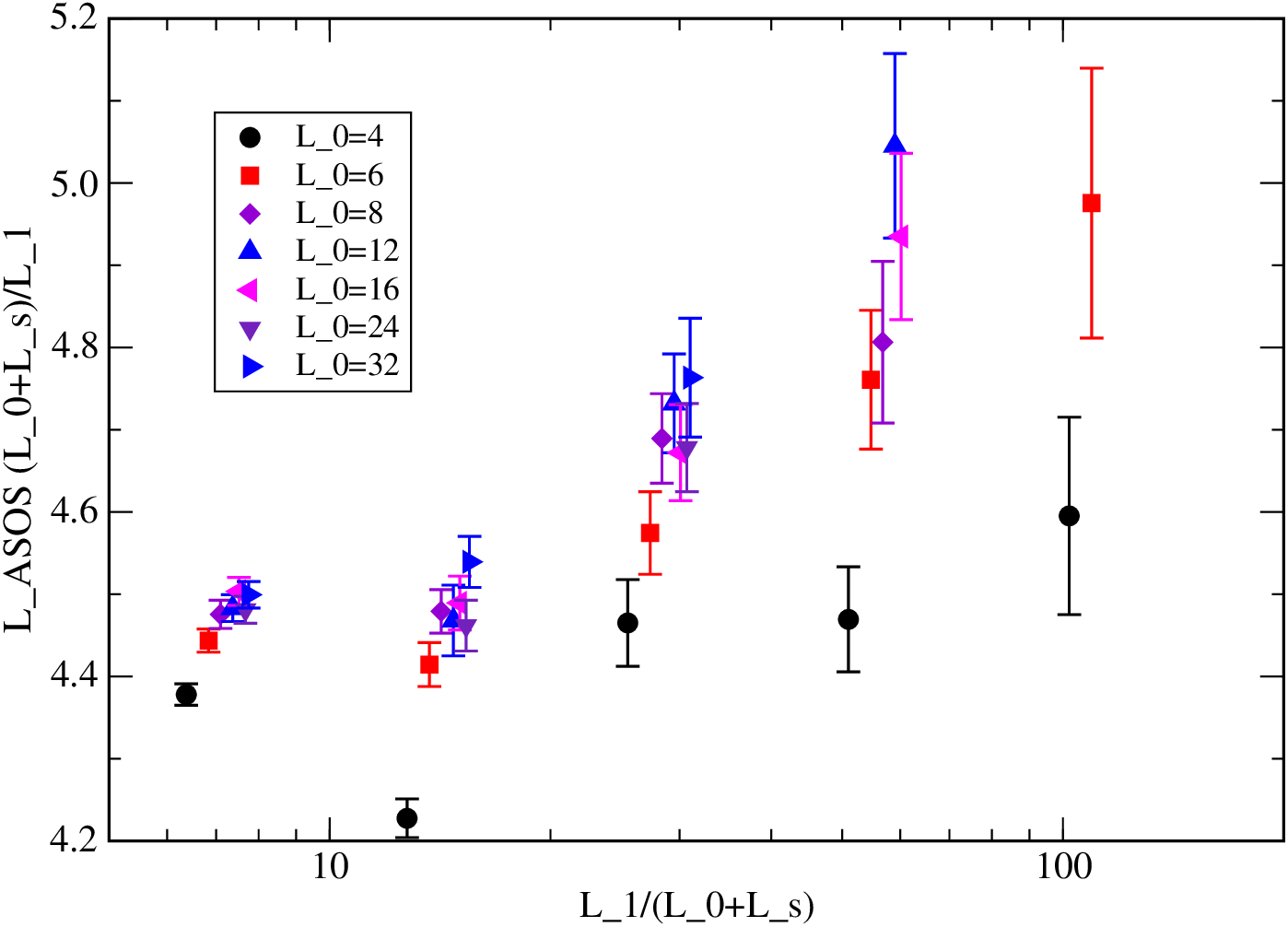}
}
\end{center}
\caption{
\label{matchingscale}  In the figure we give the matching factor
$(L_0+L_s) L_{ASOS}/L_1$ between thin films and the dual of the 
ASOS model.
}
\end{figure}

In figure \ref{matchingscale} we have plotted the matching factor 
\begin{equation}
b_{ASOS,film}= L_{ASOS} (L_0+L_s)/L_1
\end{equation}
between the dual of the ASOS model and thin films,
where we have used $L_s=1.02$, eq.~(\ref{LSFSS}). 
Starting from $L_0 = 6$ the results for the same $L_1/L_0$ 
are consistent among different $L_0$, confirming the RG prediction.
The results for the scale factor are all in the range
$ 4.2 < b_{ASOS,film} < 5.2$. Unfortunately, the statistical error
increases a lot with increasing $L_1/L_0$.  Also the value of the scale factor 
$b_{ASOS,film}$ is still increasing for our largest values of $L_1/L_0$.
Taking into account this trend, we shall assume 
\begin{equation}
\label{bfinal}
b_{ASOS,film} =5.0(5)
\end{equation}
as estimate for the limit $L_1/L_0 \rightarrow \infty$ in the following.

Using the matching factors between the dual of the ASOS model 
and other XY models that are given in ref.  \cite{HaPi97}, we arrive at
\begin{eqnarray}
b_{XY,film} = L_{XY} L_0 /L_1      &=& 1.7(2) \nonumber \\
b_{Villain,film} = L_{Villain} L_0 /L_1 &=& 0.58(6)  \nonumber \\
b_{BCSOS,film} = L_{BCSOS} L_0 /L_1   &=& 1.8(2) \;\;.
\end{eqnarray}

\begin{figure}
\begin{center}
\scalebox{0.52}
{
\includegraphics{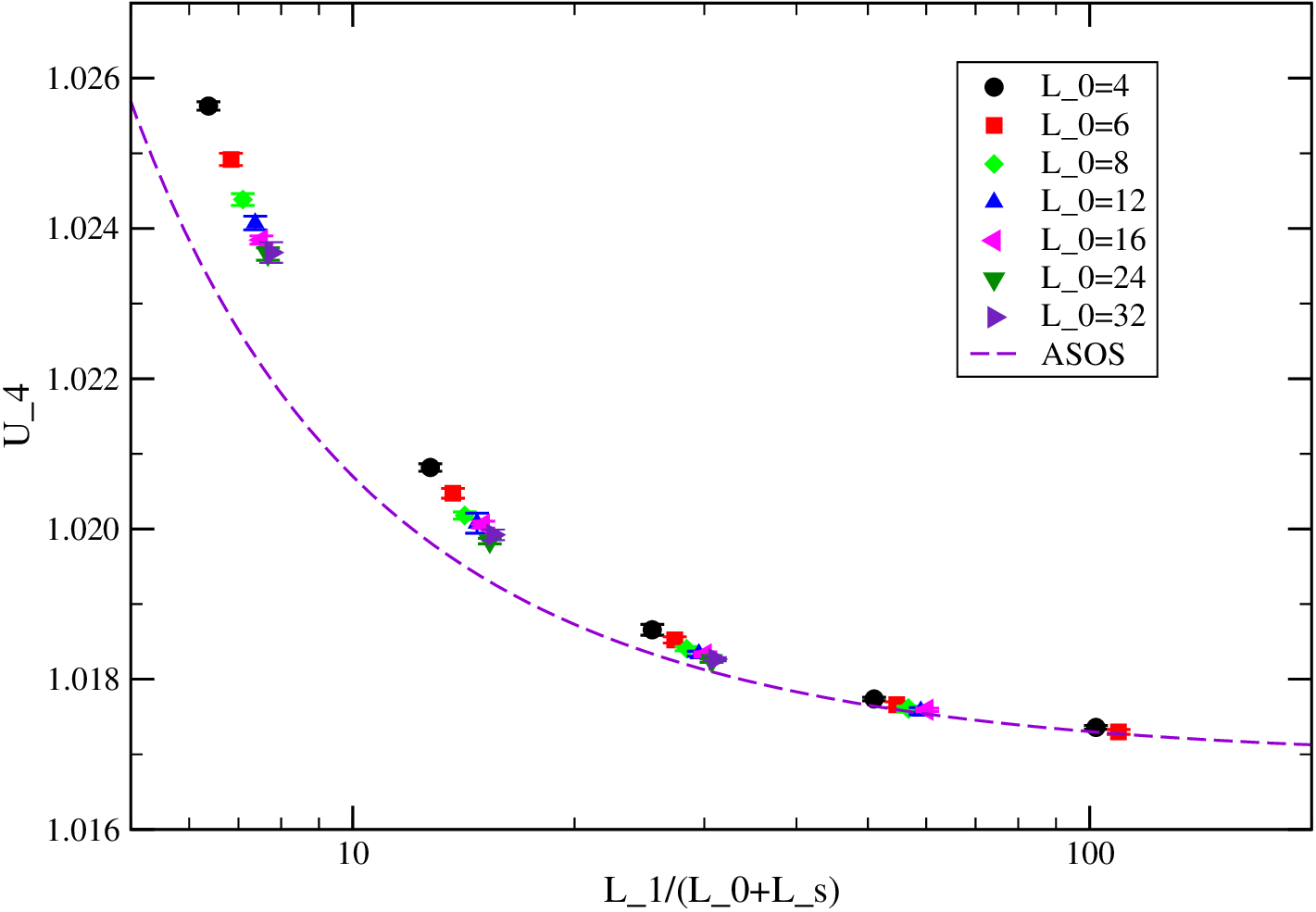}
}
\vskip1.6cm
\scalebox{0.52}
{
\includegraphics{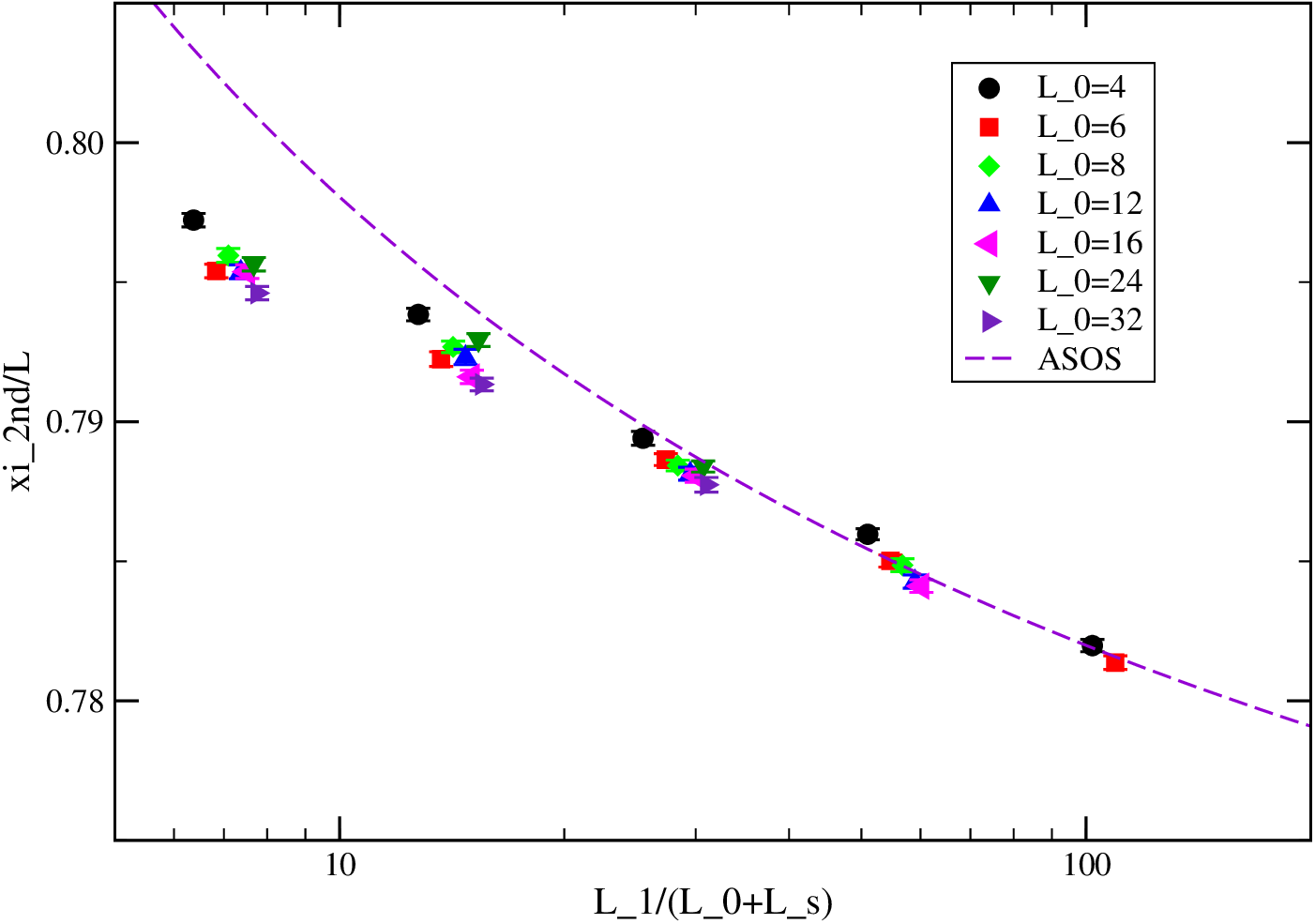}
}
\end{center}
\caption{
\label{FSSux} 
Effectively two-dimensional finite size scaling of $U_4$ 
(upper part of the figure) and 
$\xi_{2nd}/L$ (lower part of the figure). For a discussion see the text.
}
\end{figure}

In the following we like to further check that the thin films indeed undergo 
a Kosterlitz-Thouless transition. 
In figure \ref{FSSux} we give $U_4$ and $\xi_{2nd}/L$  at the estimates of $\beta_{KT}$
given in table \ref{betaKTfinal} as a function of $L_1/(L_0+L_s)$, 
where we use $L_s=1.02$, eq.~(\ref{LSFSS}).
For comparison we also give the results
for the dual of the ASOS model represented by eqs.~(\ref{Amatch80608}). 
In this case, $U_4$ and $\xi_{2nd}/L$ are plotted versus 
$L_{ASOS}/b_{ASOS,film}$, using $b_{ASOS,film}=5$.
The data for thin films with different $L_0$ fall nicely on top of each other.  
This is even true for our smallest values of $L_1/L_0$. 
The thin film results do match nicely with those of the ASOS model only for 
the larger values of $L_1/L_0$. The approach of the thin film results to those 
of the ASOS model seems compatible with power law corrections.
This behaviour is, of cause, directly related with the behaviour of 
$\beta_{KT}(L_1)$ obtained from the matching of $U_4$ and $\xi_{2nd}/L$.
Also there we have seen sizable corrections for small values of $L_1/L_0$ 
that could be fitted by a power law.

\begin{figure}
\begin{center}
\scalebox{0.52}
{
\includegraphics{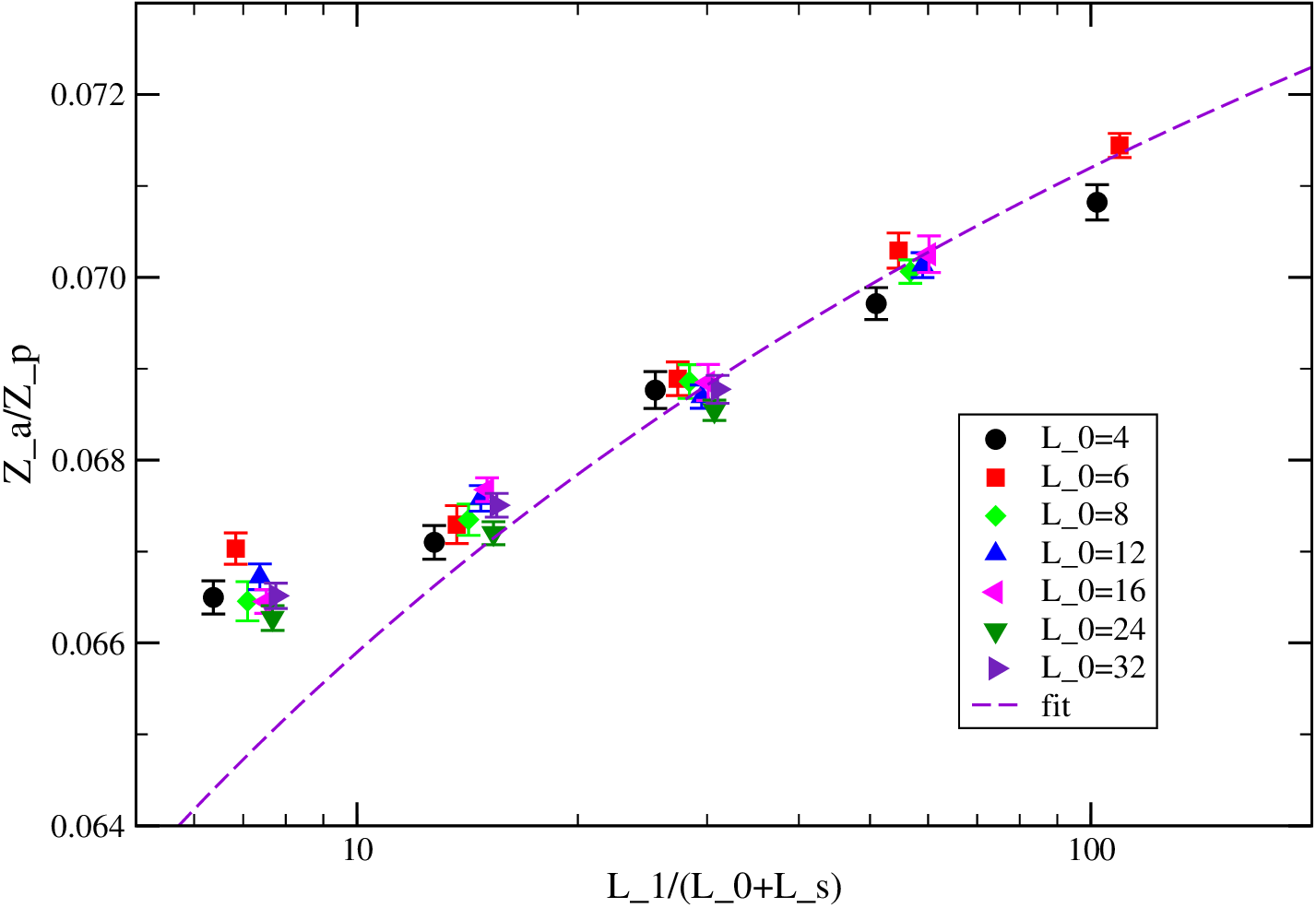}
}
\end{center}
\vskip1.1cm
\begin{center}
\scalebox{0.52}
{
\includegraphics{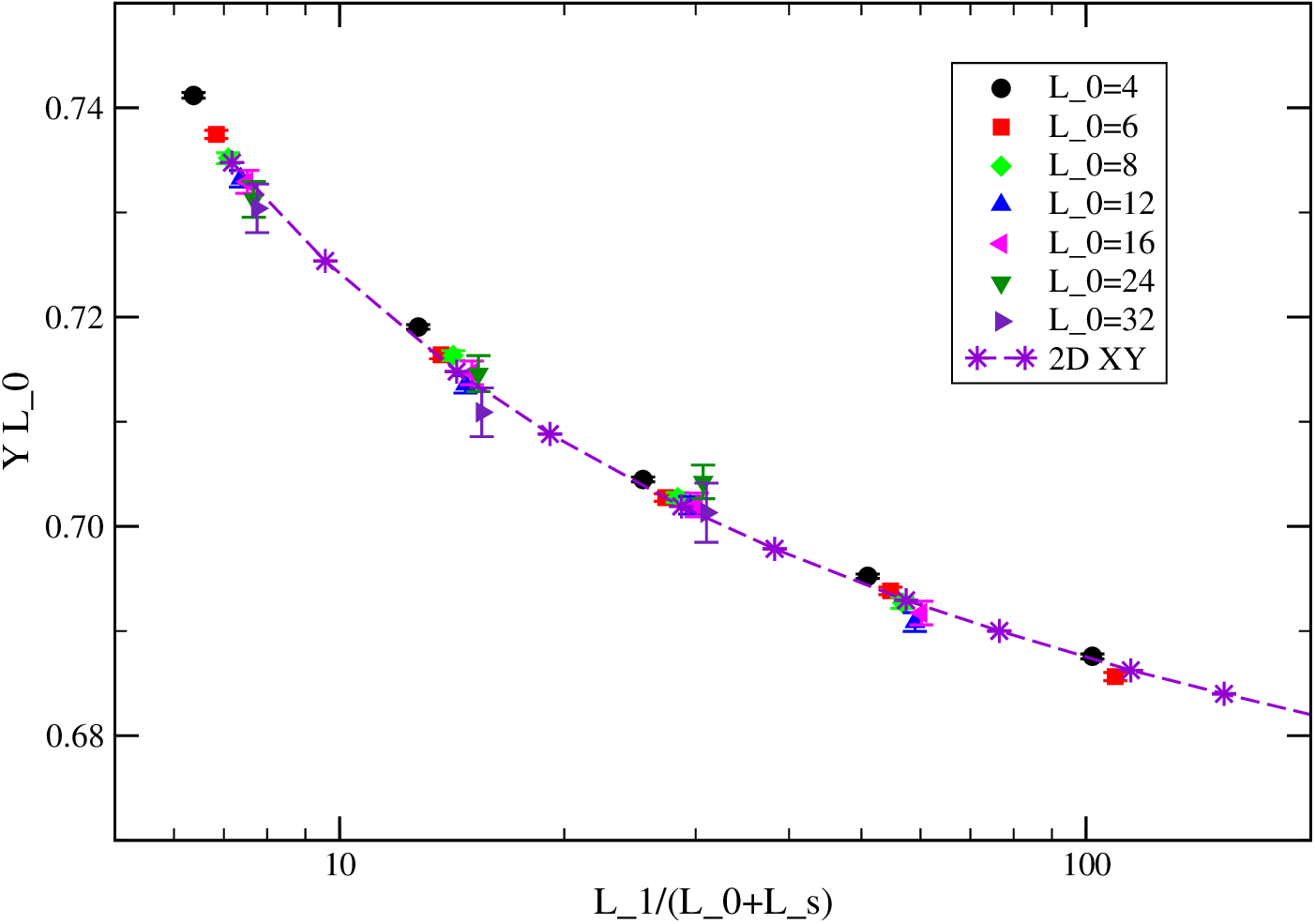}
}
\end{center}
\caption{
\label{FSSheli}
Effectively two-dimensional finite size  scaling of $Z_a/Z_p$ 
(upper part of the figure) and 
of $L_0 \Upsilon$ (lower part of the figure). For a discussion see the text
}
\end{figure}

In figure \ref{FSSheli} we give our results for the ratio of partition functions
$Z_a/Z_p$ and the helicity modulus $\Upsilon$.  In contrast to $\xi_{2nd}/L$ and 
$U_4$ these quantities have not been used to determine $\beta_{KT}$.  Therefore
they provide an additional check of the reliability of our results for 
$\beta_{KT}$ and the matching factor $b_{ASOS,film}$.
Also in the case of $Z_a/Z_p$ the data for different values of $L_0$ fall 
reasonably well on top of each other.  Unfortunately we have no data for 
$Z_a/Z_p$ for a two-dimensional XY model available. Therefore we can only 
compare  with the leading behaviour, eq.~(\ref{zazpKT}), at the KT 
transition that we derive in the appendix. We determined the constant $C$ 
by a 1-parameter fit of our data for $L_0=16$ using $L_1=512$ and $L_1=1024$,
where we have used $L=L_1/(L_0+L_s)$. The result is $C=4.31(7)$. 
The result of this fit is given in the upper part of 
figure \ref{FSSheli} by the dashed purple line.

In the lower part of figure \ref{FSSheli} we give $L_0 \Upsilon$. Note 
that, in contrast to section \ref{universal},
the $\Upsilon$ used here is measured on the thin films.
The results for different $L_0$  fall nicely on top
of each other. As expected, the relative error of $L_0 \Upsilon$ increases
with increasing thickness $L_0$ of the film.
For comparison we also plot results obtained for the 2D XY model 
\cite{myrecent}.
We have plotted results for $L_{XY}=12,16,24,32,48,64,96,128,192$ and 
$256$ as a function of $L=L_{XY}/b_{XY,film}$.
The dashed line only connects the points to guide the eye.
Here we find an almost perfect matching between the thin films and the 
two-dimensional XY model, 
starting from our smallest values of $L_1/L_0$. This suggests that corrections 
to the universal KT behaviour are very small in the helicity modulus.

\subsection{Scaling of $\beta_{KT}$ with $L_0$}
\begin{figure}
\begin{center}
\scalebox{0.52}
{
\includegraphics{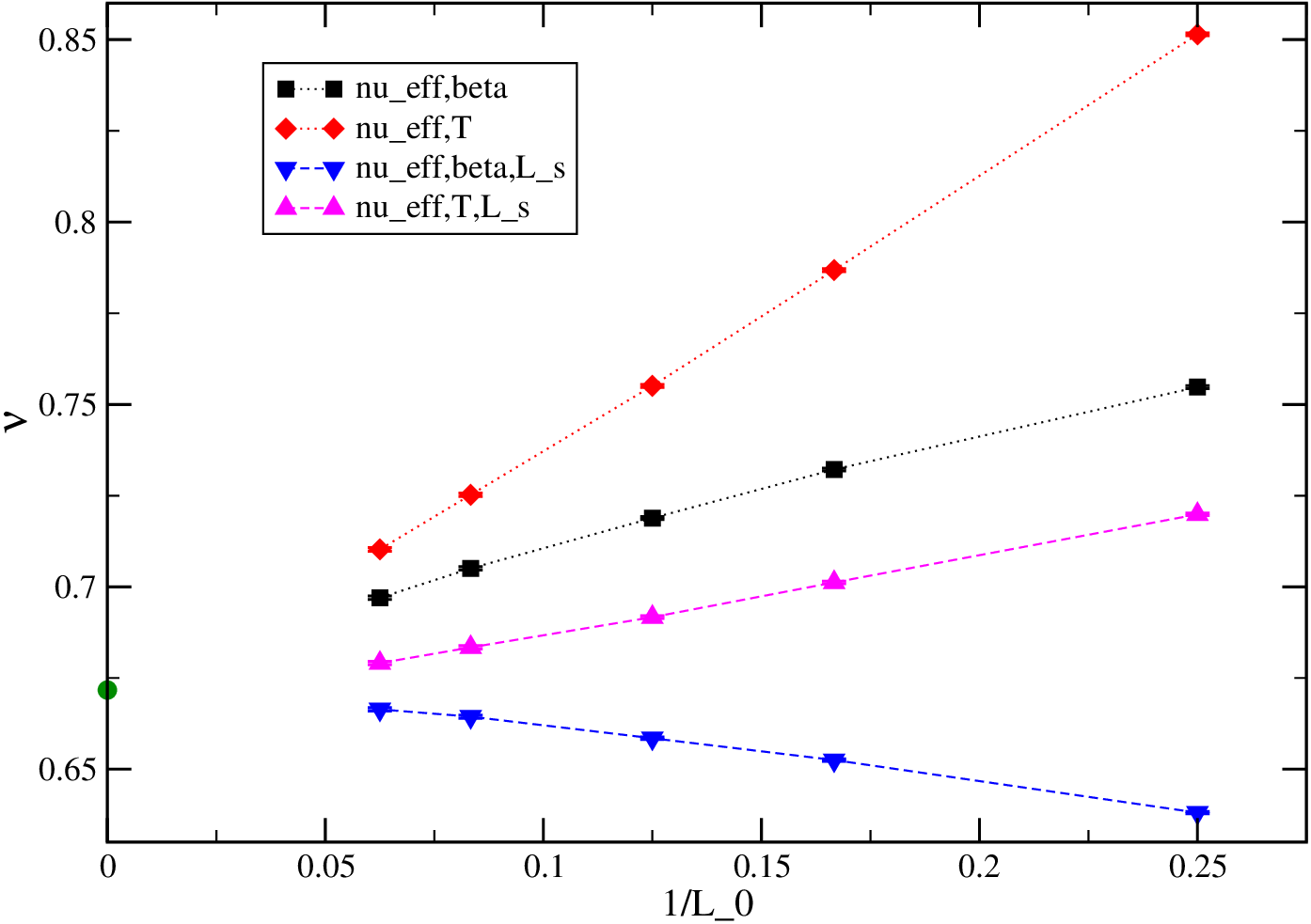}
}
\end{center}
\caption{
\label{nueffplot}
We plot the results for $\nu_{eff,\beta}(L_0)$, $\nu_{eff,T}(L_0)$,
$\nu_{eff,\beta,L_s}(L_0)$ and $\nu_{eff,T,L_s}(L_0)$  as defined  by
eqs.~(\ref{Dnueff},\ref{DnueffLs}) and the text below these equations.
The green dot gives the most precise estimate $\nu = 0.6717(1)$
\cite{recentXY}
for the critical exponent of the correlation length of the 3D XY universality 
class. The dotted lines should only guide the eye.
}
\end{figure}

As the thickness $L_0$ increases, the KT transition 
temperature of the thin film
approaches the transition temperature of the three-dimensional system. Finite
size scaling predicts that
this approach is governed by eq.~(\ref{TKTscaling}), or equivalently in terms 
of the inverse temperature
\begin{equation}
\label{betaKTscaling}
 \beta_{KT}(L_0) -  \beta_{c,3d} \simeq  L_0^{-1/\nu}  \;\; .
\end{equation}
Indeed from table \ref{betaKTfinal} we read off that $\beta_{KT}$ is
approaching $\beta_{c,3d}$ as $L_0$ is increasing. In order to check whether 
this approach is  compatible with a power law and in particular with the 
predicted exponent, we have computed
\begin{equation}
\label{Dnueff}
 \nu_{eff,\beta}(L_0)  = - \ln(2)/
  \ln\left(\frac{\beta_{KT}(2 L_0)-\beta_{c,3d}}
 {\beta_{KT}(L_0)-\beta_{c,3d}}\right) \;\;,
 \end{equation}
where we use $\beta_{c,3d}=0.5091503(6)$ \cite{recentXY}. Analogously we 
define $\nu_{eff,T}(L_0)$, where in eq.~(\ref{Dnueff}) 
$\beta_{KT}-\beta_{c,3d}$ is replaced by $T_{KT}-T_{c,3d}$. In the limit
$L_0 \rightarrow \infty$ both definitions give, by construction, 
the same result. Our results based on
the numerical estimates of $\beta_{KT}$ given in table \ref{betaKTfinal} are
plotted in figure \ref{nueffplot}.  Note that the error-bars are smaller than
the symbols. For simplicity we have added the systematic and the statistical 
error that is given  in table \ref{betaKTfinal}.
Also the error due to the uncertainty of $\beta_{c,3d}$ is taken into account.

Our results for 
$\nu_{eff,\beta}(L_0)$ and $\nu_{eff,T}(L_0)$ differ quite a lot,
indicating that analytic corrections are still important for the values 
of $L_0$ considered.
In both cases the effective exponent is decreasing with increasing $L_0$. 
Even for the largest value of $L_0$, the effective exponent is by about $4\%$ and 
$6\%$ larger than the asymptotically expected $\nu=0.6717(1)$.

In the case of the improved model, leading corrections to scaling  
are due to the free boundary conditions. 
To take these into account we  define an improved effective exponent as
\begin{equation}
\label{DnueffLs}
 \nu_{eff,\beta,L_s}(L_0)  = - \ln\left(\frac{2 L_0 + L_s}{L_0 + L_s} \right)/
   \ln\left(\frac{\beta_{KT}(2 L_0)-\beta_{c,3d}}
    {\beta_{KT}(L_0)-\beta_{c,3d}}\right) \;\;.
 \end{equation}
Also here we define $\nu_{eff,T,L_s}(L_0)$ by replacing 
$\beta_{KT}-\beta_{c,3d}$ by $T_{KT}-T_{c,3d}$ in eq.~(\ref{DnueffLs}).
Using $L_s=1.02$, eq.~(\ref{LSFSS}),
we arrive at the results that are plotted in figure \ref{nueffplot}.
Now, $\nu_{eff,\beta,L_s}$ is increasing with increasing $L_0$,
while $\nu_{eff,T,L_s}$ is decreasing. Now, for the largest value of $L_0$, 
the difference with $\nu =0.6717(1)$ is only about $1\%$ in both cases.
The remaining difference between $\nu_{eff,\beta,L_s}$ and $\nu_{eff,T,L_s}$
indicates that analytic corrections have to be taken into account.
Therefore we have fitted our data for $\beta_{KT}$ with the ansatz
\begin{equation}
\label{betaKTansatz}
 \beta_{KT}(L_0) -  \beta_{c,3D}  = a (L_0 +L_s)^{-1/\nu}  \;\;
 \times (1 + c (L_0 +L_s)^{-1/\nu}) \;\;,
\end{equation}
where we have included the leading analytic correction in addition to 
the correction caused by the boundaries.  We take $a$, $L_s$ and $c$ 
as free parameters of the fit, while now
$\beta_{c,3D}=0.5091503$ and $\nu=0.6717$ are fixed. I.e.
here we assume the RG prediction for the exponent to be correct. 
We have not included corrections $\propto L^{-\omega}$ since for 
$\lambda=2.1$ the amplitude of these corrections should be small. Furthermore 
it should be difficult to disentangle them from the boundary corrections
which are $\propto L_0^{-1}$.
The results of these fits are given in table \ref{fitbetaKT}.
\begin{table}
\caption{\sl \label{fitbetaKT}
Results of fits with the ansatz~(\ref{betaKTansatz}). $L_{0,min}$ is the 
minimal thickness that is taken into account. 
As discussed in the text, $a$, $L_s$ and $c$ are the parameters of the fit.
}
\begin{center}
\begin{tabular}{|c|c|c|c|c|}
\hline
$L_{0,min}$ &      $a$      &     $L_s$   &  $c$ & $\chi^2/$d.o.f. \\
\hline
    4       &  1.0321(11)  &    1.158(29)& 1.38(11) & 3.80 \\
    6       &  1.0286(13)  &    1.037(36)& 0.90(13) & 0.39 \\
    8       &  1.0284(18)  &    1.030(59)& 0.87(23) & 0.57 \\
\hline
\end{tabular}
\end{center}
\end{table}
Next we have repeated the fit, replacing $\beta$ by $T$ in the 
ansatz~(\ref{betaKTansatz}):
\begin{equation}
\label{betaKTansatzT}
T_{c,3D} - T_{KT}(L_0) = \tilde a (L_0 +L_s)^{-1/\nu}
 \times (1 + \tilde c (L_0 +L_s)^{-1/\nu})  \;\;.
\end{equation}

\begin{table}
\caption{\sl \label{fitTKT}
Similar to table \ref{fitbetaKT} but with the ansatz~(\ref{betaKTansatzT}).
}
\begin{center}
\begin{tabular}{|c|c|c|c|c|}
\hline
$L_{0,min}$ &      $\tilde a$      &     $L_s$   &  $\tilde c$ & $\chi^2/$d.o.f. \\
\hline
  4         &  3.9788(25)     & 1.118(11)   & -0.851(24)& 2.43  \\
  6         &  3.9697(35)     & 1.057(21)   & -1.03(6)\phantom{00}  & 0.43 \\
  8         &  3.9684(55)     & 1.047(37)   & -1.07(11)\phantom{0} & 0.60 \\
\hline
\end{tabular}
\end{center}
\end{table}

For both types of fits we find $\chi^2/$d.o.f. $< 1$ already for 
$L_{0,min}=6$, where we have included all data with $L_0\ge L_{0,min}$ 
into the fit. The result for $L_s$ from the fits with $L_{0,min}=6$ and $8$ 
are compatible among each other. Furthermore they are in perfect 
agreement with the result $L_s=1.02(7)$, eq.~(\ref{LSFSS}), obtained from
FSS at $\beta_{c,3d}$.
Also the results for the leading amplitude are compatible among the two
types of fits:
Using $a= \tilde a \beta_c^2$  we convert the results of $\tilde a$ given 
in table \ref{fitTKT} to $a=1.0314(6)$, $1.0291(9)$ and 
$1.0287(14)$ for $L_{0,min}=4$, $6$ and $8$, respectively.
I.e. the difference between the two types of fits is negligible.  

We take our final result for the amplitude from the fit with 
ansatz~(\ref{betaKTansatz}) and $L_{0,min}=8$.
We have repeated this fit for shifted values of the input parameters:
for $\beta_{c,3d}=0.5091509$, $\nu=0.6717$ and for $\beta_{c,3d}=0.5091503$, 
$\nu=0.6718$. From the results of these fits we conclude for the amplitude
\begin{equation}
\label{aresult}
 a = 1.028(2) -10.5 \times (\nu-0.6717)  -400 \times (\beta_{c,3d} - 0.5091503)\;\;.
\end{equation}

The analysis of the behaviour of $\beta_{KT}$ presented in this section
leaves little doubt that the finite size scaling 
prediction~(\ref{TKTscaling},\ref{betaKTscaling}) is correct. To fit data 
for thicknesses in the range $6 \le L_0 \le 32$ 
properly, boundary corrections and leading analytic corrections have 
to be taken into account. Note that in the case of models which are not 
improved, also corrections $\propto L_0^{-\omega}$ should  play a role.

\subsection{Universal amplitude ratios}
\label{universal}
As we have noted already in the introduction, finite size scaling predicts
that the KT transition should occur at a universal value of $L_0/\xi$,
where $\xi$ is a correlation length of the three-dimensional bulk system.
In this section we shall compute this ratio, using the transversal correlation
length $\xi_{\perp}=1/\Upsilon$ and the second 
moment correlation length $\xi_{2nd}$ in the high temperature phase.

For the following discussion it is more convenient to consider the inverse
of the function $\beta_{KT} (L_0)$, namely the thickness $L_{0,KT}(\beta)$
where the KT transition occurs at a given value of $\beta$. 
Of course this is not well defined, since $L_0$ can take only
integer values, and therefore for almost all values of $\beta$ there is no 
such  $L_0$. However by proper interpolation this problem can be solved, 
at least on a practical level. 

There are several strategies to compute the universal amplitude ratios
from Monte Carlo data. For a discussion see e.g. ref. \cite{CaHa97}.
If both quantities live in the same phase, as it is the case
here for $L_{0,KT}$ and $\Upsilon$, we can simply consider the product at
the same value of $\beta$, or equivalently the same temperature $T$.  
Then the critical limit is taken
\begin{equation}  
\label{defineLY}
 [L_{0,KT}  \Upsilon]^* \equiv
\lim_{\beta \rightarrow \beta_{c,3d}}  L_{0,KT} (\beta) \Upsilon(\beta) \;\;.
\end{equation}
One expects from RG theory that the approach to the critical limit is 
described by
\begin{equation}
\label{ampcorrections}
 (L_{0,KT} + L_s) \Upsilon = [L_{0,KT}  \Upsilon]^*  + c L_{0,KT}^{-\omega} + ...\;.
\end{equation}
If the two quantities live in different phases, as it is the case here for 
$L_{0,KT}$ and  $\xi_{2nd}$ in the high temperature phase,  we define
\begin{equation}
\label{defineLxi}
 [L_{0,KT}/\xi_{2nd}]^*  \equiv
\lim_{\beta \rightarrow \beta_{c,3d}} 
 L_{0,KT} (\beta)/\xi_{2nd}(2 \beta_{c,3d}-\beta) \;\;.
\end{equation}
I.e. the two quantities are taken at the same distance from the critical 
point of the three-dimensional system. This is however not uniquely defined. 
Instead of the same distance in $\beta$ we could also use the same 
distance in $T$. Therefore, in contrast to eq.~(\ref{ampcorrections}) also
analytic corrections appear
\begin{equation}
\label{ampcorrections2}
 (L_{0,KT}(\beta) + L_s)/\xi_{2nd}(2 \beta_{c,3d}-\beta) = 
 [L_{0,KT}/\xi_{2nd}]^*  + c L_{0,KT}^{-\omega} + ...+ d L_{0,KT}^{-1/\nu} + ... \;.
\end{equation}
The universal ratio can also be expressed in terms of amplitudes:
\begin{equation}
[L_{0,KT}/\xi_{2nd}]^* =  a^{\nu} f_{2nd,+}^{-1} \;\;,
\end{equation}
where $a$ is defined by eq.~(\ref{betaKTansatz}) and $f_{2nd,+}$ is the 
amplitude of the second moment correlation length in the high 
temperature phase.

\subsubsection{The second moment correlation length in the high 
temperature phase}
Here we analyse the Monte Carlo data for $\xi_{2nd}$ given in table 5 of 
ref. \cite{myamplitude}.
We have fitted these data with the ansatz
\begin{equation}
\label{xi2nd_ansatz}
\xi_{2nd}=f_{2nd,+} (\beta_{c,3d} -\beta)^{-\nu} \times [1 + b (\beta_{c,3d} -\beta)^{\Delta} 
 + c (\beta_{c,3d} -\beta)] \;\;,
\end{equation}
where we have included leading corrections $\propto (\beta_c -\beta)^{\Delta}$ 
and analytic corrections.  We have fixed $\nu=0.6717$, $\beta=0.5091503$
and  $\Delta = \nu \omega = 0.527$, given in ref. \cite{recentXY}.  
The results of these fits are summarized in table \ref{xi2nd_fit}.
\begin{table}
\caption{\sl \label{xi2nd_fit}
Fitting $\xi_{2nd}$ in the high temperature phase of the three-dimensional
$\phi^4$ model at $\lambda=2.1$ with the ansatz~(\ref{xi2nd_ansatz}). 
All data with $\beta \ge \beta_{min}$ are taken into account.
}
\begin{center}
\begin{tabular}{|c|c|c|c|c|}
\hline
$\beta_{min}$  & $f_{2nd,+}$ &  $b$ & $c$   & $\chi^2/$d.o.f. \\
\hline
 0.48\phantom{0}& 0.26352(5) & 0.050(4) & --0.77(2) & 1.54 \\
 0.485          & 0.26359(6) & 0.043(5) & --0.73(2) & 1.14 \\
 0.49\phantom{0}& 0.26362(8) & 0.039(8) & --0.72(4) & 1.23 \\
\hline
\end{tabular}
\end{center}
\end{table}
A $\chi^2/$d.o.f. close to one is reached for $\beta_{min}=0.485$. 
The result for
 $f_{2nd,+}$ shows very little dependence on $\beta_{min}$. 
Therefore we take as our final result the one for $\beta_{min}=0.49$:
\begin{equation}
\label{fresult}
f_{2nd,+} = 0.26362(8) + 223 \times (\beta_{c,3d} - 0.5091503) 
 -2.1 \times (\nu-0.6717)  \;\;,
\end{equation}
where we have obtained the dependence on $\beta_{c,3d}$ and on $\nu$ by redoing 
the fits with slightly changed values of $\beta_{c,3d}$ and $\nu$. 

Now we can compute the universal ratio by using the 
results~(\ref{aresult},\ref{fresult}):
\begin{equation}
\label{amp1}
 [L_{0,KT}/\xi_{2nd}]^*  =  a^{\nu}   f_{2nd,+}^{-1}
 = 3.864(6) - 4300 (\beta_{c,3d}-0.5091503) + 5 (\nu-0.6717) \;\;.
\end{equation}
Note that the error of this product is mainly due to the error of $a$.
Taking into account the uncertainty in $\beta_c$ and $\nu$ we arrive at
\begin{equation}
\label{amp2}
[L_{0,KT}/\xi_{2nd}]^*   = 3.864(9) \;\;.
\end{equation}
One should note that the exponential correlation length $\xi_{exp}$,
which describes the asymptotic decay of the correlation function,
differs only slightly from $\xi_{2nd}$ which has been used here. Following
ref. \cite{ourXY}:
\begin{equation}
 \lim_{\beta \rightarrow \beta_{c,3d}} \frac{\xi_{exp}}{\xi_{2nd}} = 1.000204(3)  \;\;,\;\;\;\; (\beta<\beta_{c,3d})
 \;\;.
\end{equation}
Hence at the level of our accuracy this difference plays no role.

\subsubsection{The helicity modulus in the low temperature phase}
The results for the helicity modulus, which are summarized 
in the second column of table \ref{tableL0KTU}, are taken from ref. 
\cite{myamplitude}.
In order to obtain $L_{0,KT}(\beta)$ at the values of $\beta$ that 
were simulated in ref. \cite{myamplitude}, we have to interpolate the results 
given in table \ref{betaKTfinal}. To this end we have used  the 
ansatz~(\ref{betaKTansatz}) along with the values for the coefficients
obtained from the fit with $L_{0,min}=8$. The numbers for $L_{0,KT}(\beta)$
that are given in the third column of table \ref{tableL0KTU} are obtained 
by numerically inverting eq.~(\ref{betaKTansatz}).
Note that for the present purpose, the asymptotic correctness of the 
coefficients of eq.~(\ref{betaKTansatz}) is less important.
For the present purpose it is sufficient that the function accurately describes
our data in the  range of $L_0$ that has been simulated.
In table \ref{tableL0KTU} we quote no error for $L_{0,KT}$. Given the accuracy 
of $\beta_{KT}(L_0)$ we expect an relative error of about 1/2 \textperthousand
\phantom{x} for $L_{0,KT}$ at a fixed value of $\beta$.

\begin{figure}
\begin{center}
\scalebox{0.52}
{
\includegraphics{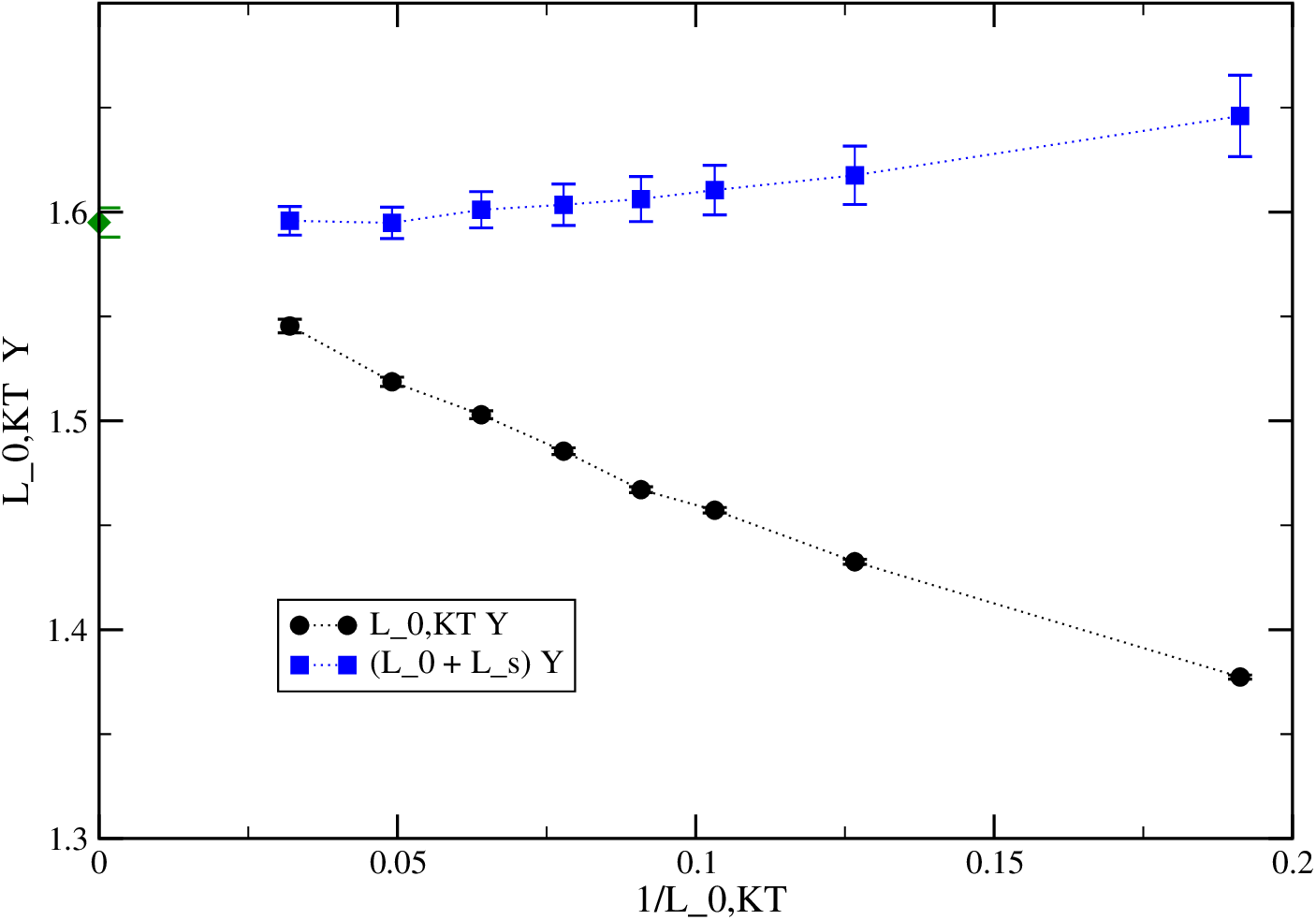}
}
\end{center}
\caption{
\label{LYplot}
The numerical estimates of $L_{0,KT} \Upsilon$ and 
$(L_{0,KT}+L_s)  \Upsilon$  which are quoted in table \ref{tableL0KTU}
are plotted as a function of $1/L_{0,KT}$. In  addition we give our final 
result~(\ref{UYfinal}) (green diamond) 
for the limit $L_{0,KT} \rightarrow \infty$.
For a discussion see the text.
}
\end{figure}

In the fourth column of table \ref{tableL0KTU} we have computed the product
$ L_{0,KT}(\beta) \Upsilon(\beta)$; in () we give  the error due to the error of 
$\Upsilon$ and in $[]$ the 1/2 \textperthousand \phantom{x} error assumed 
for $L_{0,KT}(\beta)$. 
$ L_{0,KT}(\beta) \Upsilon(\beta)$ is increasing with decreasing $\beta$. Even 
for the two smallest values of the inverse temperature $\beta=0.52$ and 
$\beta=0.515$, the difference of $ L_{0,KT}(\beta) \Upsilon(\beta) $ is larger than 
the error. In order to take into account the boundary corrections, which 
are leading for the improved model, we have computed $(L_{0,KT}+L_s) \Upsilon$.
The results are given in the last column of table \ref{tableL0KTU}. We have used 
the numerical value $L_s=1.02(7)$, eq.~(\ref{LSFSS}). The error of $L_s$ is reflected 
by the number given in $\{\}$.  Now the dependence of the product on $\beta$ is much 
reduced. Starting from $\beta=0.53$, the results are consistent within the 
quoted errors.
\begin{table}
\caption{\sl \label{tableL0KTU}
Results for the product $L_{0,KT}(\beta) \Upsilon(\beta)$ 
}
\begin{center}
\begin{tabular}{|l|l|l|l|l|}
\hline
\multicolumn{1}{|c}{$\beta$} &
\multicolumn{1}{|c}{$\Upsilon$} &
\multicolumn{1}{|c}{$L_{0,KT}$} &
\multicolumn{1}{|c}{$ L_{0,KT} \Upsilon$} &
\multicolumn{1}{|c|}{$(L_{0,KT}+L_s)  \Upsilon$} \\
\hline
  0.515 & 0.04939(8)  &           31.29 &  1.5454(25)[8]  &  1.5958(26)[8]$\{35\}$ \\
  0.52  & 0.07456(7)  &           20.37 &  1.5187(14)[8]  &  1.5948(15)[8]$\{52\}$ \\
  0.525 & 0.09628(7)  &           15.61 &  1.5029(11)[8]  &  1.6011(12)[8]$\{67\}$ \\
  0.53  & 0.11565(7)  &          12.845 &  1.4855(9)[7]   &  1.6035(10)[8]$\{81\}$ \\
  0.535 & 0.13349(6)  &          11.012 &  1.4670(7)[7]   &  1.6062(7)[8]$\{93\}$  \\
  0.54  & 0.15035(6)  &\phantom{0}9.692 &  1.4572(6)[7]   &  1.6105(6)[8]$\{105\}$ \\
  0.55  & 0.18145(6)  &\phantom{0}7.895 &  1.4325(5)[7]   &  1.6176(5)[8]$\{127\}$ \\
  0.58  & 0.26340(5)  &\phantom{0}5.229 &  1.3773(3)[7]   &  1.6460(3)[8]$\{184\}$ \\
\hline
\end{tabular}
\end{center}
\end{table}
As our final result we take 
\begin{equation}
\label{UYfinal}
[L_{0,KT} \Upsilon]^* = 1.595(7)
\end{equation}
which is the average of the result for $\beta = 0.515$ and $0.52$. The error-bar
includes all results up to $\beta=0.53$. 
For illustration we have  plotted in fig. \ref{LYplot} the results given table 
\ref{tableL0KTU} along with our final result~(\ref{UYfinal}). 

\section{Preliminary results for the dynamically diluted XY model and the 
standard XY model}
We have performed a preliminary study of the dynamically diluted XY model at
$D=1.02$ and 
the standard XY model.  To this end we have simulated lattices of the size 
$L_1=L_2=16 \times L_0$ with $L_0=4,8,16$ and 32. In order to determine 
$\beta_{KT}$ we assume that effectively two-dimensional finite size effects
are universal in 
thin films in the 3D XY universality class. I.e. that they are the same 
as for the $\phi^4$ model studied in section \ref{KTphi4}. In particular 
we read off from figure \ref{FSSux}  that $\xi_{2nd}/L=0.792(2)$ for 
$L_1/L_0=16$ at the KT transition temperature. 
In the following we have determined an estimate of $\beta_{KT}$
by requiring  that $\xi_{2nd}/L$ assumes the value $0.792$ on lattices
with $L_1/L_0=16$. The results
obtained this way are summarized in table \ref{prelimddXY} for the 
dynamically diluted XY model at $D=1.02$ 
and in table  \ref{prelimXY} for the standard
XY model. The number given in $()$ is the statistical error, while 
the number given $[]$ is the error due to the uncertainty of 
$\xi_{2nd}/L$. 

As a consistency check we have computed $L_0 \Upsilon$ and  $U_4$
at the values of $\beta_{KT}$ summarized in tables 
\ref{prelimddXY} and \ref{prelimXY}. Indeed we find that $U_4 \approx 1.02$
and $L_0 \Upsilon  \approx 0.715$ as we also read off for $L_1/L_0=16$ 
from figures \ref{FSSux} and \ref{FSSheli} for the $\phi^4$ model, 
respectively.

\begin{table}
\caption{\sl \label{prelimddXY}
Preliminary estimates of $\beta_{KT}$ for the dynamically diluted XY model
at $D=1.02$.
The results are all obtained from lattices of the size $L_1=L_2=16 \times L_0$
by requiring $\xi_{2nd}/L=0.792(2)$. For a discussion see the text.
}
\begin{center}
\begin{tabular}{|r|l|l|}
\hline
 $L_0$     &  $\beta_{KT}$  \\
\hline
 4 &  0.6767(3)[2] \\
 8 &  0.60871(14)[6] \\
16 &  0.58083(11)[3] \\
32 &  0.57012(3)[2]  \\
\hline
\end{tabular}
\end{center}
\end{table}

The results for the  ddXY model, including all data, can be fitted by
\begin{equation}
\label{betaKTddXY}
\mbox{ddXY}, D=1.02: \;\;\;\;\; 
\beta_{KT} =0.5637963 +   1.1286(71) \times (L_0 + 0.70(3))^{-1/0.6717} \;.
\end{equation}
Since we have only data for a few thicknesses of the film, 
we cannot check for systematic errors due to subleading corrections.

In table  \ref{prelimXY}, for the standard XY model we also give the results 
that can be found in table I of ref. \cite{SchMa97} which were obtained with 
staggered boundary conditions. 
Therefore no exact match with our results is expected.
For $L_0=4$ the inverse KT transition temperature 
$\beta_{KT}$ of ref. \cite{SchMa97} is clearly smaller than ours.
For $L_0=8$ it is still slightly smaller, while for $L_0=16$ it is slightly 
larger than ours.

\begin{table}
\caption{\sl \label{prelimXY}
Preliminary results for $\beta_{KT}$ for the standard XY model are
summarized in the second column. In the third column we give
for comparison the results obtained in ref. \cite{SchMa97}.
Note that these refer to staggered boundary conditions and therefore 
no exact match of the numbers is expected.
}
\begin{center}
\begin{tabular}{|r|l|l|}
\hline
 $L_0$     &  $\beta_{KT}$, ours &  $\beta_{KT}$, \cite{SchMa97}  \\
 \hline
  4 &  0.5665(3)[3] & 0.5448(3) \\
  8 &  0.4987(2)[1] & 0.4949(10)\\
 12 &               & 0.47934(23)\\
 16 &  0.47132(3)[3]& 0.47226(7)\\
 20 &               & 0.46847(15)\\
 32 &  0.46050(2)[1]& \\
\hline
\end{tabular}
\end{center}
\end{table}

Fitting the results for the standard XY model is more difficult than 
those of the ddXY model at $D=1.02$, since for the  standard XY model
we expect a sizable correction $\propto L^{-\omega}$ with 
$\omega \approx 0.785(20)$ \cite{recentXY}.
Ignoring these corrections we get from fitting all our data of table 
\ref{prelimXY} 
\begin{equation}
\label{betaKTXY}
\mbox{XY}: \;\;\;\;\;
\beta_{KT} =0.4541655 +   1.1387(45) \times (L_0 + 0.75(2))^{-1/0.6717} 
\phantom{xxxxxxxxxx}
\end{equation}
with $\chi^2/$d.o.f. $=2.02$.  Taking into account the correction
$\propto L^{-\omega}$ instead of the boundary correction we obtain
\begin{equation}
\label{betaKTXY2}
\mbox{XY}: \;\;\;\;\;
\beta_{KT} =0.4541655 + 1.1545(46) \times (1 - 0.70(2) \times L_0^{-0.785}) 
\times L_0^{-1/0.6717} 
\end{equation}
with  $\chi^2/$d.o.f. $=1.92$. Given the small number of data points 
available
we made no attempt to fit both types of corrections simultaneously.
The difference between the two fits~(\ref{betaKTXY},\ref{betaKTXY2}) might
give some indication of systematic errors.

Fitting the values of  $\beta_{KT}$ for $L_0=8,12,16$ and $20$
of ref. \cite{SchMa97} we get
\begin{equation}
 \beta_{KT,staggered} =0.4541655   + 1.47(13) 
 \times (L_0 + 3.13(76))^{-1/0.6717} \phantom{xxxxxxxxxxxx}
\end{equation}
or alternatively, taking into account corrections $\propto L_0^{-\omega}$:
\begin{equation}
\beta_{KT,staggered} =0.4541655 + 1.44(10) \times (1 - 1.92(25) \times L_0^{-0.785})
\times L_0^{-1/0.6717} \;.
\end{equation}
If free and staggered boundary conditions are equivalent, the prefactor 
of $L_0^{-1/0.6717}$ should be identical. Apparently, here we find a 
discrepancy of about three times the error. Based on the data of 
ref. \cite{SchMa97} it is hard to decide whether this discrepancy is due to 
subleading corrections which are not taken into account or a subtle 
cancellation of corrections $\propto L_0^{-\omega}$ and
$\propto L_0^{-1}$ or that free and staggered boundary conditions are, 
against our expectation, not equivalent.

In relation with ref. \cite{recentXY} we have also determined the 
correlation length of the standard XY model and the ddXY at $D=1.02$ in the 
high temperature phase.  Using a similar analysis as for the $\phi^4$ model
at $\lambda=2.1$ in the previous section, we arrive at
$f_{2nd,+} = 0.2822(3)$ for the ddXY model at $D=1.02$ and 
$f_{2nd,+} = 0.2880(3)$
for the standard XY model, where the amplitude $f_{2nd,+}$ 
is defined by eq.~(\ref{xi2nd_ansatz}). With eq.~(\ref{amp1})  we arrive at
$[L_{0,KT}/\xi_{2nd}]^*=3.84(2)$ using eq.~(\ref{betaKTddXY}) 
for the ddXY model at $D=1.02$ 
and $[L_{0,KT}/\xi_{2nd}]^*=3.79(1)$ using eq.~(\ref{betaKTXY}) and
 $[L_{0,KT}/\xi_{2nd}]^*=3.82(1)$ using eq.~(\ref{betaKTXY2}) for the 
standard XY model.  Given the fact that systematic errors are not 
fully taken into account, we regard these results as consistent 
with our more precise estimate~(\ref{amp2}) obtained from the 
$\phi^4$ model at $\lambda=2.1$, confirming universality. 

\section{Comparison with experimental results}
To our knowledge, there are no previous theoretical results for the amplitude 
ratios $[L_{0,KT} \Upsilon]^*$ and $[L_{0,KT}/\xi_{2nd}]^*$. 

The experimental 
study of $^4$He provides no access to the correlation length in the high temperature
phase. On the other hand the helicity modulus can be computed from the superfluid
density.
Following ref. \cite{FiBaJa73} the helicity modulus as defined here is given by
\begin{equation}
\label{super}
\Upsilon(T) = \frac{\hbar^2 \rho_{sb}(T)}{m^2 k_b T} \;\;,
\end{equation}
where $\rho_{sb}(T)$ is the superfluid density of the bulk system and $m$
the mass of a $^4$He  atom. In the literature the inverse of the helicity 
modulus is called transverse correlation length $\xi_T=1/\Upsilon$. 
The superfluid density can be obtained from measurements
of the specific heat and the velocity of the second sound.

A number of experimental studies of the KT transition in thin films of $^4$He have 
been performed. For an overview see ref. \cite{GaKiMoDi08}. 
Only a few of these works give results for the 
KT temperature, obtained e.g. from the onset of the superfluidity, for a sufficiently 
large range of the thickness of the film to extract a result for the universal 
combination $[L_{0,KT} \Upsilon]^*$.  In fact in two of them an explicit  result
for $[L_{0,KT} \Upsilon]^*$ is quoted:

Sabisky and Anderson \cite{SaAn73} have performed experiments with films of $^4$He 
on various substrates. They have studied films of a thickness up to about $75$ \AA. 
For films on CaF$_2$ they find for $L_0 \gtrapprox 35$ \AA   $\;$ 
power law scaling of the film 
thickness with the reduced KT temperature with an exponent $2/3$, 
which is in reasonable agreement with the correlation length exponent $\nu=0.6717(1)$
\cite{recentXY}.
Concerning the scaling of the thickness of the film with the superfluid density of 
the bulk system they write "the product $(\rho_s/\rho) \xi(T) = 5.8 T$ 
over the entire temperature range". In their notation $\xi(T)$ is the thickness of 
the film where the KT transition takes place at the temperature $T$.
It is clear from the text that Sabisky and Anderson have omitted a factor
\AA$/$K on the right hand side of the equation. Plugging in  numerical values  
for $\rho$, $\hbar$, $m$ and $T_{\lambda}$ into eq.~(\ref{super}) we arrive at
\begin{equation}
[L_{0,KT} \Upsilon]^* = 5.8 \times 0.2664 =  1.545  \;\;.
\end{equation}

The authors of ref. \cite{Laar95} have studied Helium films on glass of a thickness 
up to $47$ \AA. They also find power law scaling of the KT thickness of the film 
with the temperature. Their numerical estimate for the exponent is $0.71$.
Even more interesting, they quote in their eq.~(7):
\begin{equation}
 \delta_c(T) = 0.43 + 1.61 \xi_{\perp}(T) \;\;,
\end{equation}
where $\delta_c(T)$ is the KT thickness of the film  at the temperature $T$ and 
$\xi_{\perp}(T)$ is the transverse correlation length of the bulk. 
I.e. they give a direct 
estimate for the universal combination which we are interested in. It  
matches perfectly with our theoretical result.

It is a bit puzzling that there are more recent experimental studies using larger 
thicknesses of the $^4$He film, which fit less well the theoretical expectation
than the two reported above.
Yu et al. \cite{YuFiGa89} have studied $^4$He films on Kapton up to a thickness 
of $156$ \AA.  They see a power law scaling, starting from $25 $ \AA. 
However they find an exponent $0.52(1)$ which is clearly smaller than $\nu=0.6717(1)$.
Furthermore, from their figure 3  it can be clearly seen that the ratio of 
$\xi_{\perp}(T)$ and the KT thickness varies a lot over the range of 
thicknesses that has been 
studied. For their largest thickness, $156$ \AA $\;$  we read off from their figure
$L_{0,KT}/\xi_{\perp}  \approx 10^{0.2} \approx 1.58$. On the other hand, for 
$25 $ \AA $\;$ one reads off $L_{0,KT}/\xi_{\perp} \approx 10^{0.4} 
\approx 2.51$. The reduced transition temperatures are $0.2541$ and $0.0073$
for  $25 $ \AA $\;$ and $156$ \AA  $\;$, respectively \cite{Kimball}.

In refs. \cite{GaMe98,KiGa01} the KT temperature has been determined 
for films of the thickness  $483$ \AA $\;$ and $2113$ \AA  $\;$ 
at saturated vapour pressure. These films
are confined by two wafers of silicon. The  adiabatic fountain resonance
method has been used to determine the KT transition temperature.
The reduced transition temperatures are 
$\approx 1.29 \times 10^{-3}$ and $\approx 1.64 \times 10^{-4}$ for 
$483$ \AA $\;$ and $2113$ \AA  $\;$, respectively  \cite{Kimball}.
An estimate of 
$\xi_{\perp}$ can be obtained e.g. from  ref. \cite{SiAh84}: from their 
figure 3 we read off $\xi_0 \approx 3.42$ \AA  $\;$  at saturated vapour
pressure, where $\xi_{\perp} \simeq \xi_0 t^{-\nu}$. It follows 
$L_{0,KT}/\xi_{\perp} \approx 1.62$ and $1.77$ 
for $483$ \AA $\;$ and $2113$ \AA  $\;$, respectively.  

\section{Summary and Conclusions}
We have simulated lattice models in the three-dimensional XY universality 
class with thin film geometry. In order
to mimic the vanishing order parameter at the boundaries that is observed
in experiments on thin films of $^4$He we have implemented  free boundary 
conditions. These boundary conditions lead to corrections to finite size scaling 
$\propto L_0^{-1}$, where $L_0$ is the thickness of the film.
These corrections can be described by $L_{0,eff} =  L_0 + L_s$.  Furthermore  
one expects corrections to finite size scaling $\propto L_0^{-\omega}$ 
\cite{Barber}, where $\omega=0.785(20)$ \cite{recentXY}. 
Analysing Monte Carlo data, it is  difficult to disentangle corrections
with exponents that are so close. In fact in previous studies, 
e.g. \cite{SchMa97},  
of thin films, using the standard XY model, only the boundary corrections have 
been taken into account, which might lead to systematic errors in the 
extrapolation to the critical limit.
In order to avoid this problem, we have studied improved models. In 
particular we have studied in detail the $\phi^4$ model on a simple
cubic lattice at $\lambda=2.1$. This value of the parameter $\lambda$
is close to the most recent estimate $\lambda^*=2.15(5)$ \cite{recentXY}, 
where $\lambda^*$ is defined such that the amplitude of leading 
corrections exactly vanishes. At  $\lambda=2.1$ the amplitude of corrections
to scaling should be at least by a factor of 20 smaller than in the 
standard XY model.

First we have studied the $\phi^4$ model at the critical point
of three-dimensional system. 
We  have simulated lattices  of the size $L_0=L_1=L_2$ and $L_0+1=L_1=L_2$
with $L_1=8,12,16,24,32$ and 48. We employed free boundary conditions in 
0-direction, while periodic boundary conditions are employed in 
1 and 2-direction. We have measured the four phenomenological couplings
$U_4$, $U_6$, $\xi_{2nd}/L$ and $Z_a/Z_p$. The latter two quantities 
are taken only for the 1 and the 2-direction.
We clearly see corrections $\propto L_0^{-1}$ in all of these quantities.
This is in clear contrast with simulations, where periodic boundary
conditions in all directions are employed. E.g. in ref. \cite{recentXY},
using improved models one sees corrections $\propto L_0^{-\epsilon}$, where
$1.6  \lessapprox \epsilon  \lessapprox 2$.
Casting the corrections $\propto L_0^{-1}$ into the form $L_{0,eff}=L_0+L_s$, 
one gets, within the error, the {\sl same} value for $L_s$ from
all four quantities. As our final result we quote $L_s=1.02(7)$. Note 
that $L_s$ depends on the model and on the particular boundary 
conditions that are used. In future studies of other properties of thin 
films, like e.g. the thermodynamic Casimir force, using the same lattice 
model as here, the knowledge of $L_s$ is a valuable asset.

Next we studied the phase transition in  thin films.
To this end, we applied a finite size scaling method that we
have proposed recently \cite{myrecent}.  Finite size scaling of the 
effectively two-dimensional film means $L_0 \ll L_1,L_2 \lessapprox \xi$, where 
$\xi$ is the correlation length in the 1  or 2-direction of a system with the 
thickness $L_0$ and infinite extension in the other two directions.
Also here, we have studied in detail the $\phi^4$ model at $\lambda=2.1$.
We have simulated lattices up to $L_1/L_0 =128$ for $L_0 =4$ and $6$, up to 
$L_1/L_0 =64$ for $L_0 =8$, $12$ and $16$ and up to $L_1/L_0 =32$ for $L_0 =24$
and $32$. Matching the finite size behaviour of the second moment correlation length
$\xi_{2nd}/L$ and the Binder cumulant $U_4$ with that of the dual
of the ASOS model at the Kosterlitz-Thouless (KT) transition we obtain 
accurate estimates of the
transition temperature of the thin films. Note that the finite size scaling 
behaviour is very similar for different two-dimensional XY models 
\cite{myrecent}.
The dual of the ASOS model has been chosen in ref. \cite{myrecent} as reference 
model for purely technical reason.
Furthermore, we obtain an estimate for the scale factor $b_{ASOS,film}$
that relates the effective two-dimensional lattice size  $L_1/L_0$ of the 
thin film and that of the dual of the ASOS model.
The finite size scaling behaviour of $Z_a/Z_p$ and the helicity modulus times
the thickness $L_0 \Upsilon$, which were not used to determine $\beta_{KT}$
and $b_{ASOS,film}$, clearly confirm the KT nature of the transition in the thin films.

Next we have studied the scaling of the inverse of the KT temperature 
$\beta_{KT}$ with the thickness 
$L_0$ of the film.  We find that for the  thicknesses that we have 
studied, even in the improved model, corrections to scaling are important.
Using the ansatz~(\ref{betaKTansatz}) that includes  corrections due to the 
free boundary conditions as well as leading analytic corrections we can fit,
fixing $\nu=0.6717(1)$, all data with $L_0 \ge 6$, with $\chi^2/$d.o.f.=0.39.

For the first time, we provide a theoretical estimate for the amplitude ratio 
\begin{equation}
 [L_{0,KT}/\xi_{\perp}]^*  = 1.595(7) \;\;,
\end{equation}
where $\xi_{\perp}$ is the transversal correlation length in the low temperature
phase of the three-dimensional bulk system. Note that $\xi_{\perp}$ is the inverse of 
the helicity modulus as defined here.  This result nicely compares with the 
experimental results of refs. \cite{SaAn73,Laar95}.  On the other hand, more
recent experiments \cite{GaMe98,KiGa01} on $^4$He films with thicker films seem 
to match less well.

We have also computed the universal amplitude ratio 
$[L_{0,KT}/\xi_{2nd}]^*  =3.864(9)$, where $\xi_{2nd}$ is the second moment 
correlation length of the three-dimensional bulk system in the high temperature 
phase.
We also provide some preliminary results for the KT transition temperature for 
the dynamically diluted XY model at $D=1.02$ and the standard XY model for 
$L_0=4,8,16$ and $32$. Using the data of these two models we get results
for $[L_{0,KT}/\xi_{2nd}]^*$ that are essentially consistent with the one quoted 
above for the $\phi^4$ model, confirming the universality of the result.

\section{Appendix A: Ratio of partition functions $Z_a/Z_p$ at the KT transition}
In the spin-wave approximation of a two-dimensional XY model, 
the partition function is given by
\begin{equation}
 Z_{SW} = \sum_{n_1,n_2} W(n_1,n_2) Z(0,0) \;\;,
\end{equation}
where $n_1$ and $n_2$ count the windings of the XY field along the 
1 and 2 directions. In the case $L_1=L_2$, for isotropic couplings, 
the weights are given by
\begin{equation}
 W(n_1,n_2) 
 = \exp\left(-2 \pi^2  \beta_{SW} [n_1^2 + n_2^2]\right)  \;\;,
\end{equation}
where for periodic boundary conditions, $n_1$ and $n_2$ take 
integer values and in the case of antiperiodic boundary conditions
half-integer. I.e. in the spin-wave approximation
\begin{equation}
 \frac{Z_a}{Z_p} = 
 \frac{\sum_{n_1=-\infty}^{\infty} \sum_{n_2=-\infty}^{\infty}  
  \exp\left(-2 \pi^2  \beta_{SW} [(n_1+1/2)^2 + n_2^2]\right)}
 {\sum_{n_1=-\infty}^{\infty} \sum_{n_2=-\infty}^{\infty}
 \exp\left(-2 \pi^2  \beta_{SW} [n_1^2 + n_2^2]\right)} \;\;.
 \end{equation}

The KT transition is characterized by $\beta_{SW} = 2/\pi$. For 
the neighbourhood of the KT transition we get
\begin{equation}
Z_a/Z_p = 0.0864272337... - 0.426489404... (\beta-2/\pi) + ... \;\;.
\end{equation}
At the KT transition the $\beta_{SW}$ depends on the scale as
\begin{equation}
\beta_{SW}= 2/\pi + \frac{1}{\pi} \frac{1}{\ln L +C} + ... \;\;,
\end{equation}
where in finite size scaling, we identify this scale with the lattice size.
It follows for lattices with $L_1=L_2=L$:
\begin{equation}
\label{zazpKT}
Z_a/Z_p = 0.0864272337... - 0.135755793... \frac{1}{\ln L +C} + ... \;\;.
\end{equation}

\section{Acknowledgements}
I like thank  W. Janke for discussions and encouragement. Thanks to
E. Vicari for reading the final version of the draft.
I am grateful to M. Kimball for remarks on the status of experiments and 
for providing me with explicit results for reduced KT transition temperatures.
This work was supported by the Deutsche Forschungsgemeinschaft (DFG) 
under grant No JA 483/23-1.


\begin{thebibliography}{99}
\bibitem{WiKo}
Wilson K G and Kogut J,
{\sl The renormalization group and the $\epsilon$-expansion}, 1974
Phys.\ Rep.\ C {\bf 12} 75

\bibitem{Fisher74}
Fisher M E, {\sl The renormalization group in the theory of critical behavior},
1974  Rev.\ Mod.\ Phys.\ {\bf 46} 597

\bibitem{Fisher98}
Fisher M E,
{\sl Renormalization group theory: 
Its basis and formulation in statistical physics}, 1998
 Rev.\ Mod.\ Phys.\ {\bf 70} 653

\bibitem{PeVi02}
Pelissetto A and Vicari E,
{\sl Critical Phenomena and Renormalization-Group Theory}, 2002
Phys.\ Rept.\ {\bf 368} 549 [cond-mat/0012164]


\bibitem{LSNCI-96}
Lipa J A, Swanson D R, Nissen J A, Chui T C P and
Israelsson U E,
{\sl
Heat Capacity and Thermal Relaxation of Bulk Helium very near the $\lambda$-Point}, 1996
Phys.\ Rev.\ Lett.\ {\bf 76} 944

\bibitem{lipa2003}
Lipa J A, Nissen J A, Stricker D A, Swanson D R and Chui T C P,
{ \sl
Specific heat of liquid helium in zero gravity very near the $\lambda$-point},
2003
Phys.\ Rev.\ B {\bf 68} 74518 [cond-mat/0310163]

\bibitem{SiAh84} 
Singasaas A and  Ahlers G,
{\sl Universality of static properties near the 
superfluid transition in $^4$He}, 1984
Phys.\ Rev.\ B {\bf 30} 5103 

\bibitem{GoAh92}
Goldner L S and Ahlers G,
{Superfluid fraction of $^4$He very close to $T_{\lambda}$}, 1992
Phys.\ Rev.\ B {\bf 45} 13129 

\bibitem{BaHaLiDu07}
Barmatz M, Hahn I, Lipa J A, and Duncan R V,
{\sl Critical phenomena in microgravity: Past, present, and future}, 2007
Rev.\ Mod.\ Phys.\ {\bf 79} 1

\bibitem{recentXY}
Campostrini M, Hasenbusch M, Pelissetto A,
and Vicari E,
{\sl
Theoretical estimates of the critical exponents of the superfluid
transition in He4 by lattice methods}, 2006 Phys.\ Rev.\ B {\bf 74} 144506
[cond-mat/0605083]

\bibitem{myAPAM}
Hasenbusch M, {\sl
The three-dimensional XY universality class:
A high precision Monte Carlo estimate of the universal amplitude ratio
$A_+/A_-$}, 2006
J.\ Stat.\ Mech.\  P08019 [cond-mat/0607189]

\bibitem{myamplitude}
Hasenbusch M, 
{A Monte Carlo study of the three-dimensional XY universality class: Universal
amplitude ratios}, [arXiv:0810.2716]  

\bibitem{KT}
Kosterlitz J M and Thouless D J,
{\sl Ordering, metastability and phase transitions in two-dimensional systems}
1973  J.\ Phys.\ C {\bf 6} 1181; 
Kosterlitz J M, {\sl The critical properties of the two-dimensional xy model},
1974 J.\ Phys.\ C {\bf 7} 1046
ne- or Two-Dimensional

\bibitem{Jo77}
Jos\'e J V, Kadanoff L P, Kirkpatrick S and Nelson D R,
{\sl Renormalization, vortices, and symmetry-breaking perturbations in the 
two-dimensional planar model}, 1977 
Phys.\ Rev.\ B {\bf 16} 1217

\bibitem{AmGoGr80}
Amit D J,  Goldschmidt Y Y and Grinstein G, 
{\sl Renormalisation group analysis of the phase transition in the 2D Coulomb
 gas, Sine-Gordon theory and XY model}, 1980
J.\ Phys.\  A {\bf 13}  585

\bibitem{Barber}
M. N. Barber {Finite-size Scaling} in {\sl Phase Transitions and Critical Phenomena, Vol. 8,}
eds. C. Domb and J. L. Lebowitz, (Academic Press, 1983)

\bibitem{Privman}
{\sl Finite Size Scaling and Numerical Simulation of Statistical Systems,}
ed. V. Privman,
(World Scientific, 1990).

\bibitem{Fi71}
M.E. Fisher, {\sl Critical Phenomena}, Proceedings of the International School 
of Physics ''Enrico Fermi´´, Varenna, Italy, Course LI, edited by M.S. Green
(Academic, New York, 1971)

\bibitem{CaFi76}
Capehart T W and  Fisher M E, 
{\sl Susceptibility scaling functions for ferromagnetic Ising films}, 1976 
Phys.\ Rev.\ B {\bf 13}  5021


\bibitem{GaKiMoDi08}
Gasparini F M, Kimball M O, Mooney K P, and Diaz-Avila M, 
{\sl Finite-size scaling of $^4$He at the superfluid transition}, 2008
Rev.\ Mod.\ Phys.\ {\bf 80} 1009

\bibitem{Dohm93}
Dohm V, {\sl The superfluid transition in confined 4 He: 
Renormalization-group theory}, 1993 
PHYSICA SCRIPTA {\bf T49A}  46.

\bibitem{NhMa03}
Nho K and Manousakis E,
{\sl Heat-capacity scaling function for confined superfluids}, 2003
Phys.\ Rev.\ B {\bf 68} 174503 [cond-mat/0305500]

\bibitem{Binder}
K. Binder, {\sl Critical behaviour at surfaces} in 
{\sl Phase Transitions and Critical Phenomena, Vol. 8,}
eds. C. Domb and J. L. Lebowitz, (Academic Press, 1983)  p. 1.

\bibitem{Diehl86}
H. W. Diehl, {Field-theoretical Approach to Critical Behaviour at Surfaces} in
{\sl Phase Transitions and Critical Phenomena},
edited by C. Domb and J.L. Lebowitz, Vol. 10 (Academic, London 1986) p. 76.

\bibitem{DiDiEi83}
Diehl H W, Dietrich S, and Eisenriegler E, 
{\sl Universality, irrelevant surface operators, and corrections to scaling 
in systems with free surfaces and defect planes}, 1983
Phys.\ Rev.\ B {\bf 27} 2937

\bibitem{RG}
Newman K E and Riedel E K,
{\sl
Critical exponents by the scaling-field method:
The isotropic N-vector model in three dimensions},
1984 Phys.\ Rev.\ B {\bf 30}  6615

\bibitem{JaNa93}
Janke W and Nather K, 
{\sl Monte Carlo simulation of dimensional crossover in the XY model}, 1993
Phys.\ Rev.\ B {\bf 48} 15807 

\bibitem{SchMa95a}
Schultka N and Manousakis E,
{\sl Crossover from Two- to Three-Dimensional Behavior in Superfluids}, 1995
Phys.\ Rev.\ B {\bf 51} 11712 [cond-mat/9406014]

\bibitem{SchMa96}
Schultka and E. Manousakis, 
{\sl Scaling of the superfluid density in superfluid films}, 1996
J.\  Low.\ Temp.\ Phys.\ {\bf 105}  3 [cond-mat/9602085]

\bibitem{SchMa97}
Schultka N and Manousakis E, 
{Boundary effects in superfluid films}, 1997
J.\ Low.\ Temp.\ Phys.\ {\bf 109} 733  [cond-mat/9702216]

\bibitem{ZhNhLa06}
Zhang C, Nho K, and Landau D P, 
{\sl Finite-size effects on the thermal resistivity of $^4$He in the 
quasi-two-dimensional geometry}, 2006 Phys.\  Rev.\ B {\bf 73} 174508

\bibitem{Hu07}
Hucht A,
{\sl Thermodynamic Casimir Effect in $^4$He Films near $T_c$: 
Monte Carlo Results}, 2007
Phys.\ Rev.\ Lett.\ {\bf 99} 185301 [arXiv:0706.3458]  

\bibitem{VaGaMaDi07} 
Vasilyev O, Gambassi A, Maciolek A, and Dietrich S,
{\sl Monte Carlo simulation results for critical Casimir forces}, 2007
Europhys.\ Lett.\ {\bf 80} 60009 [arXiv:0708.2902]

\bibitem{myrecent}
Hasenbusch M, 
{\sl The Binder Cumulant at the Kosterlitz-Thouless Transition}, 2008
J.\ Stat.\ Mech.\  P08003 [arXiv:0804.1880]

\bibitem{spain}
Ballesteros H G, Fernandez L A, Martin-Mayor V, and Munoz-Sudupe A,
{\sl Finite size effects on measures of critical exponents in d=3 O(N) models},
1996 Phys.\ Lett.\ B {\bf 387}  125 [cond-mat/9606203]

\bibitem{Bielefeld2}
Cucchieri A, Engels J, Holtmann S, Mendes T and Schulze T,
{\sl
Universal amplitude ratios from numerical studies of the
three-dimensional O(2) model},
2002 J.~Phys.\ A {\bf 35}  6517  [cond-mat/0202017]

\bibitem{Bloete05}
Deng Y, Bl\"ote  H W J,  and  Nightingale M P,
{\sl Surface and bulk transitions in three-dimensional O(n) models}, 2005
Phys.\ Rev.\ E {\bf 72} 016128 [cond-mat/0504173]

\bibitem{BiJa05}
Bittner E and Janke W,
{\sl Nature of Phase Transitions in a Generalized Complex $|psi|^4$ Model},
2005
Phys.\ Rev.\ B  {\bf 71} 024512 [cond-mat/0501468]

\bibitem{HaTo99}
Hasenbusch M and T\"or\"ok T,
{\sl High precision Monte Carlo study of the 3D XY-universality class},
1999 J.\ Phys.\ A {\bf 32} 6361 [cond-mat/9904408]

\bibitem{ourXY}
Campostrini M, Hasenbusch M, Pelissetto A, Rossi P,
and Vicari E,
{\sl Critical behavior of the three-dimensional XY universality class}, 2001
Phys.\ Rev.\ B {\bf 63} 214503 [cond-mat/0010360]

\bibitem{teli89}
Li Y-H and Teitel S,
{\sl Finite-size scaling study of the three-dimensional classical XY model},
1989 Phys.\ Rev.\ B {\bf 40}  9122

\bibitem{FiBaJa73}
Fisher M E, Barber M N, and Jasnow D,
{\sl Helicity Modulus, Superfluidity, and Scaling in Isotropic Systems}, 1973
Phys.\ Rev.\ A {\bf 8}  1111

\bibitem{Ha92}
Hasenbusch M,
{\sl Direct Monte Carlo Measurement of the Surface Tension in Ising Models},
1993  J.\ Phys.\ I (France) {\bf 3} 753 [hep-lat/9209016]

\bibitem{wolff}
Wolff U,
{\sl Collective Monte Carlo Updating for Spin Systems}, 1989
Phys.\ Rev.\ Lett.\ {\bf 62}  361

\bibitem{HaPiVi99}
Hasenbusch M, Pinn K and Vinti S, 
{\sl Critical Exponents of the 3D Ising Universality Class From Finite
Size Scaling With Standard and Improved Actions}, 1999
Phys.\ Rev.\ B {\bf 59} 11471
[hep-lat/9806012]

\bibitem{twister}
Saito M, {\sl An Application of Finite Field:
Design and Implementation of 128-bit Instruction-Based Fast 
Pseudorandom Number Generator}, PhD thesis, Dept. of Math., 
Graduate School of Science, Hiroshima University, Advisor: M. Matsumoto; 
The numerical program and a detailed description can be found at
``http://www.math.sci.hiroshima-u.ac.jp/{\~{}}m-mat/MT/SFMT/index.html"

\bibitem{WeMi88}
Weber H and Minnhagen P,
{\sl Monte Carlo determination of the critical temperature for the 
two-dimensional XY model}, 1988 
Phys.\ Rev.\ B {\bf 37}  5986

\bibitem{myxi}
Hasenbusch M, 
{\sl The two-dimensional XY model at the transition temperature:
A high precision Monte Carlo study}, 2005
J.\ Phys.\ A: Math. Gen. {\bf 38}  5869 [cond-mat/0502556]

\bibitem{HaPi97}
Hasenbusch M and Pinn K,
{\sl Computing The Roughening Transition of Ising and Solid-on-Solid Models
by BCSOS Model Matching}, 1997
J. Phys. A 30 (1997) 63 [cond-mat/9605019]

\bibitem{CaHa97}
Caselle M and Hasenbusch M,
{\sl Universal Amplitude Ratios in the 3-D Ising Model}, 1997
J.\ Phys.\ A {\bf 30} 4963 [hep-lat/9701007]

\bibitem{SaAn73}
Sabisky E S and  Anderson C H,
{\sl Onset for Superfluid flow in He$^4$ Films on a Variety of Substrates},
1973
Phys.\ Rev.\ Lett.\ {\bf 30} 1122

\bibitem{Laar95}
van de Laar R W A, van der Hoek A, van Beelen H, 
{\sl Two-dimensional behaviour of very thin $^4$He films on glass}, 1995
Physica B {\bf 216}  24.

\bibitem{YuFiGa89}
Yu Y Y, Finotello D and Gasparini F M, 
{\sl Finite-size scaling and the convective conductance and specific heat
 of planar helium films near the superfluid transition}, 1989
Phys.\ Rev.\ B {\bf 39} 6519

\bibitem{Kimball}
Kimball M O, private communication

\bibitem{GaMe98}
Gasparini F M and  Metha S
{\sl Temperature oszillations from a temperature superleak}, 1998
J.\ Low\ Temp.\ Phys.\ {\bf 110} 293

\bibitem{KiGa01}
Kimball M O and Gasparini F M,
{\sl Superfluid fraction of He-3-He-4 mixtures confined at 0.0483 $\mu$ m between
silicon wafers}, 2001
Phys.\ Rev.\ Lett.\ {\bf 86} 1558

\end{thebibliography}
\end{document}